\documentclass[12pt]{article}
\usepackage{eurosym}
\usepackage{graphicx}
\usepackage{pdfpages}
\usepackage{authblk}
\usepackage[numbers,sort&compress]{natbib}
\usepackage{xcolor}
\usepackage[utf8]{inputenc}
\usepackage{amsmath}
\usepackage{geometry}
\title{Controlled Fission and Superposition of Vector Solitons in an Integrable Model of Two-Component Bose-Einstein Condensates}

\author[1]{V. Ramesh Kumar$^{\dag}$}
\author[1]{V. Rajadurai}
\author[2,3]{Boris A. Malomed$^*$}

\affil[1]{Department of Physics, Velammal Engineering College (Autonomous), Chennai-600066, India}
\affil[2]{Department of Physical Electronics, School of Electrical Engineering, Faculty of Engineering, and Center for Light-Matter Interaction, Tel Aviv P.O. Box 39040, Israel}
\affil[3]{Instituto de Alta Investigaci\'{o}n, Universidad de Tarapac\'{a}, Casilla 7D, Arica, Chile}

\begin{document}
\maketitle

\begin{center}
Email: ramehkumar@velammal.edu.in$^{\dag}$, malomed@tauex.tau.ac.il$^{*}$
\end{center}

\begin{abstract}
We investigate the dynamics of vector solitons in a two-component
Bose-Einstein condensates governed by the system of
Gross-Pitaevskii equations. Using a gauge-transformation approach, we
construct a four-soliton solution and analyze their interactions, including
superposition states, fission, and shape-preserving collisions. We explore
the ability of time-dependent parameters, such as the intra- and
intercomponent interaction coefficients and trapping potential, to control
the soliton properties. In particular, we demonstrate controlled
four-soliton fission, highlighting its potential applications to quantum
data processing and coherent matter-wave transport. The results suggest
experimental realization in BEC systems and provide insights into nonlinear
wave interactions in multicomponent quantum fluids.
\end{abstract}
%\maketitle

\section{Introduction}

The realization of Bose-Einstein condensates (BECs) in ultracold atomic
gases \cite{bose1924,einstein1924,anderson1995,ketterle1969,Hulet} has
greatly advanced the study of quantum matter. It offers a platform for
investigating nonlinear coherent structures, including soliton, vortices,
and other wave excitations \cite%
{strecker2002,burger1999,frantzeskakis2010,theocharis2003,liang2005}. Among
these, bright solitons have attracted much interest due to their remarkable
stability and coherence in attractive- or tunable-interaction regimes. The
dynamics of bright solitons can be effectively described within the
framework of the Gross-Pitaevskii (GP) equations, which are nonlinear Schr%
\"{o}dinger equation (NLSE) adapted for dilute trapped quantum gases \cite%
{dalfovo1999, pethick2003}. In particular, quasi-one-dimensional BECs allow
for the balance between nonlinearity and gradient energy, enabling the
formation and manipulation of solitons \cite{strecker2002, liang2005}. Dark
solitons, which represent another important class of nonlinear excitations,
have also been created in BECs \cite{burger1999}. These structures exhibit
intriguing properties, including phase shifts and long lifetimes, which have
been broadly studied both theoretically and experimentally \cite%
{frantzeskakis2010}. In addition, ring dark solitons and vortex necklaces
have been proposed as stable excitations in the effectively two-dimensional
setting \cite{theocharis2003}.\newline

Multicomponent Bose-Einstein condensates (MBECs), which are composed of two
or more interacting atomic states or species, present a fundamentally more
complex dynamics compared to single-component condensates. The interaction
between intra- and interspecies forces leads to the emergence of unique
structures, such as soliton trains, domain walls, and spin-switching regimes
\cite{ho1996, Kasamatsu2005, Kevrekidis2008,Stoof2009}. A crucial feature of
these systems is the presence of vector solitons, where multiple solitonic
components coexist within various internal states or atomic species \cite%
{Ieda2004, carr2017}. These solitons have unique characteristics, including
fusion, splitting, and changes in amplitude driven by parameter adjustments
\cite{he2023}. The dynamics of solitons are significantly influenced by
external potentials, intra- and interspecies scattering lengths, as well as
other variables \cite{Busch2001, Ohberg2001, Kasamatsu2006}.

Research on solitons in BEC parallels the investigations of similar settings
in other domains of physics \cite{Dauxois}. In particular, in nonlinear
optics, bright and dark solitons, as well as interactions between them, have
been thoroughly examined in optical fibers \cite{KA}.The study of soliton
collision dynamics across different nonlinear-optical setups has advanced
the understanding of energy localization \cite{porsezian1994}. Additionally,
examining soliton interactions in the framework of coupled nonlinear Schr%
\"{o}dinger equations has uncovered the potential for shape-altering
collisions, which can be used in the design of optical logic gates \cite%
{radhakrishnan1996, kanna2001}.\newline

Earlier research primarily concentrated on interactions between two solitons
in BECs. Expanding this to include three- and four-soliton solutions
presents new challenges and possibilities. This paper explores the dynamics
of multi-soliton interactions in a two-component BEC, using integrable
coupled Gross-Pitaevskii equations. Explicit three- and four-soliton
solutions are developed by means of the gauge transformation method,
allowing the examination of controlled soliton fission, superposition
states, and elastic collisions affected by time-varying interaction
parameters and external potentials. These solutions may find applications to
quantum data processing and coherent matter-wave control. In Section 2, we
introduce the model of the two-component BEC. In Section 3, we establish the
integrability of the system, using the respective Lax pair, and derive the
associated compatibility conditions. In Section 4, we construct explicit
multi-soliton solutions, including one-, two-, three-, and four-soliton
states, using the gauge-transformation method. In Section 5, we analyze the
dynamics of the superposition four-soliton state, controlled soliton
fission, and its similarity to quantum data processing. Finally, in Section
6 we summarize our findings and suggest possible directions for the
extension of the work.

\section{The Model}

We consider a two-component BEC in which the components represent two
hyperfine states of the same atomic species. The evolution of the system is
governed by the coupled GP equations:
\begin{eqnarray}
i\hbar \frac{\partial \psi _{1}}{\partial t} &=&\left( -\frac{\hbar ^{2}}{%
2m_{1}}\nabla ^{2}+\mathcal{U}_{11}|\psi _{1}|^{2}+\mathcal{U}_{12}|\psi
_{2}|^{2}+\mathcal{V}_{1}\right) \psi _{1}, \\
i\hbar \frac{\partial \psi _{2}}{\partial t} &=&\left( -\frac{\hbar ^{2}}{%
2m_{2}}\nabla ^{2}+\mathcal{U}_{21}|\psi _{1}|^{2}+\mathcal{U}_{22}|\psi
_{2}|^{2}+\mathcal{V}_{2}\right) \psi _{2}.
\end{eqnarray}%
Here, $\psi _{1}$ and $\psi _{2}$ are the condensate wave functions,
normalized to the respective particle numbers, $\mathcal{U}_{11,22}$ and $%
\mathcal{U}_{12,21}$ being coefficients of the intra- and inter-species
interactions, respectively. The external trapping potentials $\mathcal{V}%
_{1,2}$ are taken as quadratic (harmonic-oscillator) ones, leading to the
quasi-one-dimensional description when the transverse motion is suppressed.

To consider the system in which the two components represent different
hyperfine states of the same atom, we take equal atomic masses $%
m_{1}=m_{2}=m $ and equal trapping potential in both components. By defining
appropriate dimensionless parameters and rescaling the time and coordinate
variables, the system reduces to the following coupled dimensionless GP
equations \cite{vrk2010}:
\begin{eqnarray}
i\frac{\partial \psi _{1}}{\partial t}+\frac{1}{2}\frac{\partial ^{2}\psi
_{1}}{\partial x^{2}}+(b_{11}(t)|\psi _{1}|^{2}+b_{12}(t)|\psi
_{2}|^{2})\psi _{1}+iG_{1}(t)\psi _{1}+\lambda ^{2}(t)x^{2}\psi _{1} &=&0,
\label{3} \\
i\frac{\partial \psi _{2}}{\partial t}+\frac{1}{2}\frac{\partial ^{2}\psi
_{2}}{\partial x^{2}}+(b_{21}(t)|\psi _{1}|^{2}+b_{22}(t)|\psi
_{2}|^{2})\psi _{2}+iG_{2}(t)\psi _{2}+\lambda ^{2}(t)x^{2}\psi _{2} &=&0.
\label{4}
\end{eqnarray}%
In these equations the effective interaction coefficients $b_{ij}(t)$ may be
functions of time, which may be implemented by means of the Feshbach
resonance controlled by a time-dependent spatially uniform magnetic field
\cite{Caputo,Ueda}. In addition, the equations also include the gain/loss
terms with variable coefficients $G_{1,2}(t)$, and the trapping potential
may also be time-dependent, as represented by coefficient $\lambda ^{2}(t)$.
In fact, it may assume a sign-changing form, thus switching between the
trapping and expulsive potentials \cite{Sandy}. The system simplifies to the
symmetric integrable Manakov model, which permits multi-soliton solutions
with shape-preserving collisions, assuming constant coefficients $%
b_{11}=b_{22}=b_{12}=b_{21}$ and the absence of the gain/loss and
external-potential terms, $G_{1,2}=\lambda =0$ \cite{manakov1974}.

We consider the case when the system of Eqs. (\ref{3}) and (\ref{4}) can be
reduced to an integrable form. Then, we use the gauge-transformation
approach \cite{llchau}, which makes it possible to construct higher-order
soliton solutions in a systematic way. To gain a better understanding of the
dynamics of nonlinear excitations in the two-component BEC, we show how
multi-soliton interactions can be tuned by varying the interaction
parameters and the strength of the trapping potential.

\section{The Lax pair and integrability condition}

We consider the case where the system of Eqs. (3) and (4) can be reduced to
an integrable form. Then, we use the gauge transformation approach \cite%
{llchau}, which enables us to construct higher-order soliton solutions in a
systematic way.

In the framework of the AKNS formalism (Ablowitz-Kaup-Newell-Segur), a
nonlinear integrable system can be represented as the compatibility
condition of an overdetermined linear system involving an auxiliary function
$\Phi $. Specifically, the system of equations \cite{vrk2010}:
\begin{eqnarray}
\Phi _{x} &=&Q_{1}\Phi   \label{ch6eq8} \\
\Phi _{t} &=&Q_{2}\Phi   \label{ch6eq9}
\end{eqnarray}%
where, $\Phi =(\phi _{1},\phi _{2},\phi _{3})^{T}$ and the subscript $x$ in $%
\Phi _{x}$ denotes the partial derivative $\partial \Phi /\partial x$, and
similarly for $\Phi _{t}$. These form a Lax pair whose compatibility
condition ($\Phi _{xt}=\Phi _{tx}$) ensures integrability. The $Q_{1}$ and $%
Q_{2}$ are

\begin{equation}
Q_{1}=\left(
\begin{array}{ccc}
-i\zeta (t) & U & V \\
-U^{\ast } & i\zeta (t) & 0 \\
-V^{\ast } & 0 & i\zeta (t) \\
&  &
\end{array}%
\right) ,  \label{ch6eq10}
\end{equation}%
\begin{equation}
Q_{2}=\left(
\begin{array}{ccc}
-2i\zeta ^{2}(t)+2\zeta (t)i\Gamma (t)x+ & 2\zeta (t)U+ & 2\zeta (t)V+ \\
i(|U|^{2}+|V|^{2}) & i(U_{x}+2i\Gamma (t)xU) & i(V_{x}+2i\Gamma (t)xV) \\
&  &  \\
-2\zeta (t)U^{\ast }+ & 2i\zeta ^{2}(t)- & -iVU^{\ast } \\
i(U_{x}^{\ast }-2i\Gamma (t)xU^{\ast }) & 2\zeta (t)i\Gamma (t)x-i|U|^{2} &
\\
&  &  \\
-2\zeta (t)V^{\ast }+ & -iV^{\ast }U & 2i\zeta ^{2}(t)- \\
i(V_{x}^{\ast }-2i\Gamma (t)xV^{\ast }) &  & 2i\zeta (t)\Gamma (t)x-i|V|^{2}%
\end{array}%
\right) ,  \label{ch6eq11}
\end{equation}%
where
\begin{eqnarray}
U(x,t) &=&\sqrt{\alpha (t)}e^{(-i\Gamma (t)x^{2}/2)}\psi _{1}(x,t),
\label{ch6eq12} \\
V(x,t) &=&\sqrt{\beta (t)}e^{(-i\Gamma (t)x^{2}/2)}\psi _{2}(x,t),
\label{ch6eq13}
\end{eqnarray}%
and the spectral parameter $\zeta (t)$ obeys the constraint $\zeta (t)=\mu
e^{(-2\int \Gamma (t)dt)},$ where $\mu $ is a complex constant and $\Gamma
(t)$ is an arbitrary function of time. The presentation in the form of the
Lax-pair implies that the interaction coefficients are subject to
constraints $b_{11}(t)=b_{21}(t)=\alpha (t)$ and $b_{12}(t)=b_{22}(t)=\beta
(t)$. Further, the compatibility condition $Q_{1t}-Q_{2x}+[Q_{1},Q_{2}]=0$
generates Eqs.(3) and (4) with $G_{1}(t)=\Gamma (t)+\frac{1}{2}\frac{\alpha
~^{\prime }(t)}{\alpha (t)}$, $G_{2}(t)=\Gamma (t)+\frac{1}{2}\frac{\beta
~^{\prime }(t)}{\beta (t)}$, and $\lambda ^{2}(t)=\Gamma ^{2}(t)+(\Gamma
~^{\prime }(t)/2)$, with the prime standing for $d/dt$. Finally, for both
symmetric ($\alpha (t)=\beta (t)$) and antisymmetric ($\alpha (t)=-\beta (t)$%
) cases with $G_{1}(t)=G_{2}(t)$, the integrability condition becomes

\begin{equation}
\lambda ^{2}(t)=G_{1}^{2}(t)+\frac{1}{2}\frac{\left( \alpha ^{\prime
}(t)\right) ^{2}}{\alpha ^{2}(t)}-G_{1}(t)\frac{\alpha ^{\prime }(t)}{\alpha
(t)}+\frac{1}{2}G_{1}^{\prime }(t)-\frac{1}{4}\frac{\alpha ^{\prime \prime
}(t)}{\alpha (t)}.  \label{ch6eq18}
\end{equation}

Although it can be mapped to the Manakov system under specific conditions
(e.g., constant coefficients and vanishing external terms), the present
formulation allows for dynamically tunable soliton behavior, which is not
captured in traditional models.

\section{Construction of multi-soliton solutions}

\textbf{The multi-soliton solutions are obtained by successive application
of gauge transformations to the vacuum eigenfunction. Each step introduces a
soliton via a pole and an associated projection operator. The resulting
wavefunctions $\psi _{1}^{(n)}(x,t)$ and $\psi _{2}^{(n)}(x,t)$ after $n$
transformations represent the $n$-soliton solution}. To generate
two-component soliton solutions of Eqs.(3) and (4), we first consider the
vacuum solution, ($\psi _{1}^{(0)}=\psi _{2}^{(0)}=0$), so that the
corresponding eigenvalue problem becomes
\begin{eqnarray}
\Phi _{x}^{(0)} &=&Q_{1}^{(0)}\Phi ^{(0)}  \label{ch6eq19} \\
\Phi _{t}^{(0)} &=&Q_{2}^{(0)}\Phi ^{(0)}  \label{ch6eq20}
\end{eqnarray}%
where, the vacuum state $Q_{1,2}^{(0)}$ is defined as per Eqs. (\ref{ch6eq10}%
) and (\ref{ch6eq11}), respectively. Solving the above vacuum linear
eigenvalue problem, one gets
\begin{equation}
\Phi ^{(0)}=\left(
\begin{array}{ccc}
e^{-i\zeta (t)x-2i\int \zeta ^{2}(t)dt} & 0 & 0 \\
0 & e^{i\zeta (t)x+2i\int \zeta ^{2}(t)dt} & 0 \\
0 & 0 & e^{i\zeta (t)x+2i\int \zeta ^{2}(t)dt} \\
&  &
\end{array}%
\right) .  \label{ch6eq23}
\end{equation}%
\textbf{Since $\Phi =(\phi _{1},\phi _{2},\phi _{3})^{T}$ from the above one can
write, $\phi _{1}=e^{-i\zeta (t)x-2i\int \zeta ^{2}(t)\,dt},\quad \phi
_{2}=e^{i\zeta (t)x+2i\int \zeta ^{2}(t)\,dt},\quad \phi _{3}=e^{i\zeta
(t)x+2i\int \zeta ^{2}(t)\,dt}$.} Additionally, we use a transformation
function $g(x,t)$ to gauge translate the vacuum eigenfunction $\Phi ^{(0)}$
in order to obtain
\begin{eqnarray}
Q_{1}^{(1)} &=&gQ_{1}^{(0)}g^{-1}+g_{x}g^{-1}  \label{ch6eq24} \\
Q_{2}^{(1)} &=&gQ_{2}^{(0)}g^{-1}+g_{t}g^{-1}.  \label{ch6eq25}
\end{eqnarray}%
We choose the transformation function $g(x,t)$ from the solution of the
associated Riemann problem, such that it is meromorphic in the complex $%
\zeta $ plane, as
\begin{equation}
g(x,t;\zeta )=\left[ 1+\frac{\zeta _{1}-\bar{\zeta}_{1}}{\zeta -\zeta _{1}}%
P(x,t)\right] \cdot \left(
\begin{array}{ccc}
1 & 0 & 0 \\
0 & -1 & 0 \\
0 & 0 & -1 \\
&  &
\end{array}%
\right) ,  \label{ch6eq26}
\end{equation}%
where $\zeta _{1}$ and $\bar{\zeta}_{1}=\zeta _{1}^{\ast }$ are arbitrary
complex parameters, and $g^{-1}$ is the inverse matrix of $g$; $P$ is a 3$%
\times $3 projection matrix ($P^{2}=P$) to be determined. The fact that $%
Q_{1}^{(1)}$ and $Q_{2}^{(1)}$ do not develop singularities around $\zeta
=\zeta _{1}$ and $\zeta =\bar{\zeta}_{1}$ imposes the following constraints
on $P$:
\begin{eqnarray}
P_{x} &=&(1-P)\mathcal{J}Q_{1}^{(0)}(\bar{\zeta}_{1})\mathcal{J}P-P\mathcal{J%
}Q_{1}^{(0)}(\zeta _{1})\mathcal{J}(1-P)  \label{ch6eq28} \\
P_{t} &=&(1-P)\mathcal{J}Q_{2}^{(0)}(\bar{\zeta}_{1})\mathcal{J}P-P\mathcal{J%
}Q_{2}^{(0)}(\zeta _{1})\mathcal{J}(1-P)  \label{ch6eq29}
\end{eqnarray}%
where,
\begin{equation}
\mathcal{J}=\left(
\begin{array}{ccc}
1 & 0 & 0 \\
0 & -1 & 0 \\
0 & 0 & -1 \\
&  &
\end{array}%
\right) .  \label{ch6eq30}
\end{equation}%
One can generate the projection matrix $P(x,t)$ using vacuum eigenfunction $%
\Phi ^{(0)}(x,t)$ as
\begin{equation}
P=\mathcal{J}\cdot \frac{M^{(1)}}{\mathrm{{Trace}[M^{(1)}]}}\cdot \mathcal{J}%
,  \label{ch6eq31} \\
\end{equation}%
where
\begin{equation}
M^{(1)}=\Phi ^{(0)}(x,t,\bar{\zeta}_{1})\cdot \left(
\begin{array}{ccc}
e^{2\delta _{1}}\sqrt{2} & \varepsilon _{1}^{(1)}e^{2i\chi _{1}} &
\varepsilon _{2}^{(1)}e^{2i\chi _{1}} \\
\varepsilon _{1}^{\ast (1)}e^{-2i\chi _{1}} & e^{-2\delta _{1}}/\sqrt{2} & 0
\\
\varepsilon _{2}^{\ast (1)}e^{-2i\chi _{1}} & 0 & e^{-2\delta _{1}}/\sqrt{2}
\\
&  &
\end{array}%
\right) \cdot \Phi ^{(0)}(x,t,\zeta _{1})^{-1}.  \label{ch6eq33}
\end{equation}

\textbf{$M^{(1)}$ is the Hermitian $3\times 3$ matrix. Matrix $M$
essentially acts as a projection operator that encodes the structure of the
soliton.} Determinant $M^{(1)}$ vanishes under the condition $|\varepsilon
_{1}^{(1)}|^{2}+|\varepsilon _{2}^{(1)}|^{2}$=1. By choosing $\zeta
_{1}=\zeta _{11}+i\zeta _{12}$ and $\bar{\zeta}_{1}=\zeta _{1}^{\ast }$ and
using Eq. (\ref{ch6eq23}), matrix $M^{(1)}$ can be explicitly written as
\begin{equation}
M^{(1)}=\left(
\begin{array}{ccc}
e^{-\theta _{1}}\sqrt{2} & e^{-i\xi _{1}}\varepsilon _{1}^{(1)} & e^{-i\xi
_{1}}\varepsilon _{2}^{(1)} \\
e^{i\xi _{1}}\varepsilon _{1}^{\ast (1)} & e^{\theta _{1}}/\sqrt{2} & 0 \\
e^{i\xi _{1}}\varepsilon _{2}^{\ast (1)} & 0 & e^{\theta _{1}}/\sqrt{2} \\
&  &
\end{array}%
\right) ,  \label{ch6eq35}
\end{equation}%
where, $\theta _{1}=8\int \zeta _{11}(t)\zeta _{12}(t)dt+2x\zeta
_{12}(t)-2\delta _{1}$, $\xi _{1}=4\int (\zeta _{11}^{2}(t)-\zeta
_{12}^{2}(t))dt+2x\zeta _{11}(t)-2\chi _{1}$ with $\zeta _{11}(t)=\zeta
_{110}\exp \left( -\int 2\Gamma (t)dt\right) $, $\zeta _{12}(t)=\zeta
_{120}\exp \left( -\int 2\Gamma (t)dt\right) $. \textbf{Here $\zeta _{110}$ and $%
\zeta _{120}$ are constants controls the velocity, while $\delta _{1}$
determines the amplitude and $\chi _{1}$ represents the phase of soliton.}

Then substituting Eqs.(\ref{ch6eq26}) and (\ref{ch6eq28}) in Eq.(\ref%
{ch6eq24}), we obtain
\begin{equation}
Q_1^{(1)} = \left(%
\begin{array}{ccc}
-i\zeta(t) & U^{(0)} & V^{(0)} \\
-U^{(0)*} & i \zeta(t) & 0 \\
-V^{(0)*} & 0 & i \zeta(t) \\
&  &
\end{array}%
\right)-2i(\zeta_1 - \bar{\zeta_1})\left(%
\begin{array}{ccc}
0 & \tilde{P}_{12} & \tilde{P}_{13} \\
-\tilde{P}_{21} & 0 & 0 \\
-\tilde{P}_{31} & 0 & 0 \\
&  &
\end{array}%
\right),  \label{ch6eq38}
\end{equation}
and similarly for $Q_2^{(1)}$. Thus, one can obtain the one-soliton
solution, using the gauge-transformation method, as

\begin{eqnarray}
U^{(1)} &=&U^{(0)}-2i(\zeta _{1}-\bar{\zeta _{1}})\tilde{P}_{12},
\label{ch6eq39} \\
V^{(1)} &=&V^{(0)}-2i(\zeta _{1}-\bar{\zeta _{1}})\tilde{P}_{13},
\label{ch6eq40}
\end{eqnarray}%
where $\tilde{P}_{12}$ and $\tilde{P}_{13}$ are elements of the projection
matrix.

\subsection{The one-soliton solution}

Thus the explicit form of the one-soliton solution can be written as
\begin{eqnarray}
\psi _{1}^{(1)} &=&\sqrt{\frac{2}{\alpha (t)}}\varepsilon _{1}^{(1)}\zeta
_{12}(t)\mathrm{sech}(\theta _{1})\mathrm{exp\left( i(-\xi _{1}+\Gamma (t)%
\frac{x^{2}}{2})\right) ,}  \label{ch6eq41} \\
\psi _{2}^{(1)} &=&\sqrt{\frac{2}{\beta (t)}}\varepsilon _{2}^{(1)}\zeta
_{12}(t)\mathrm{sech}(\theta _{1})\mathrm{exp\left( i(-\xi _{1}+\Gamma (t)%
\frac{x^{2}}{2})\right) ,}  \label{ch6eq42}
\end{eqnarray}%
where $\alpha (t)$ and $\beta (t)$ are time-dependent interaction
coefficients, and $\varepsilon _{1,2}$ are coupling parameters should
satisfy $|\varepsilon _{1}^{(1)}|^{2}+|\varepsilon _{2}^{(1)}|^{2}$=1. To
verify the analytical solution, we performed numerical simulations using the
split-step Fourier method (SSFM). Figure 1 confirms that the numerical
simulations are in good agreement with the analytical solution. \textbf{%
Minor deviations in the soliton peak positions originate from the numerical
discretization and boundary conditions, which do not qualitatively affect
the observed dynamics.} Next, we extend our analytical approach to construct
explicit two-, three-, and four-soliton solutions.

\begin{figure}[tbp]
\begin{center}
\includegraphics[width=0.65\linewidth]{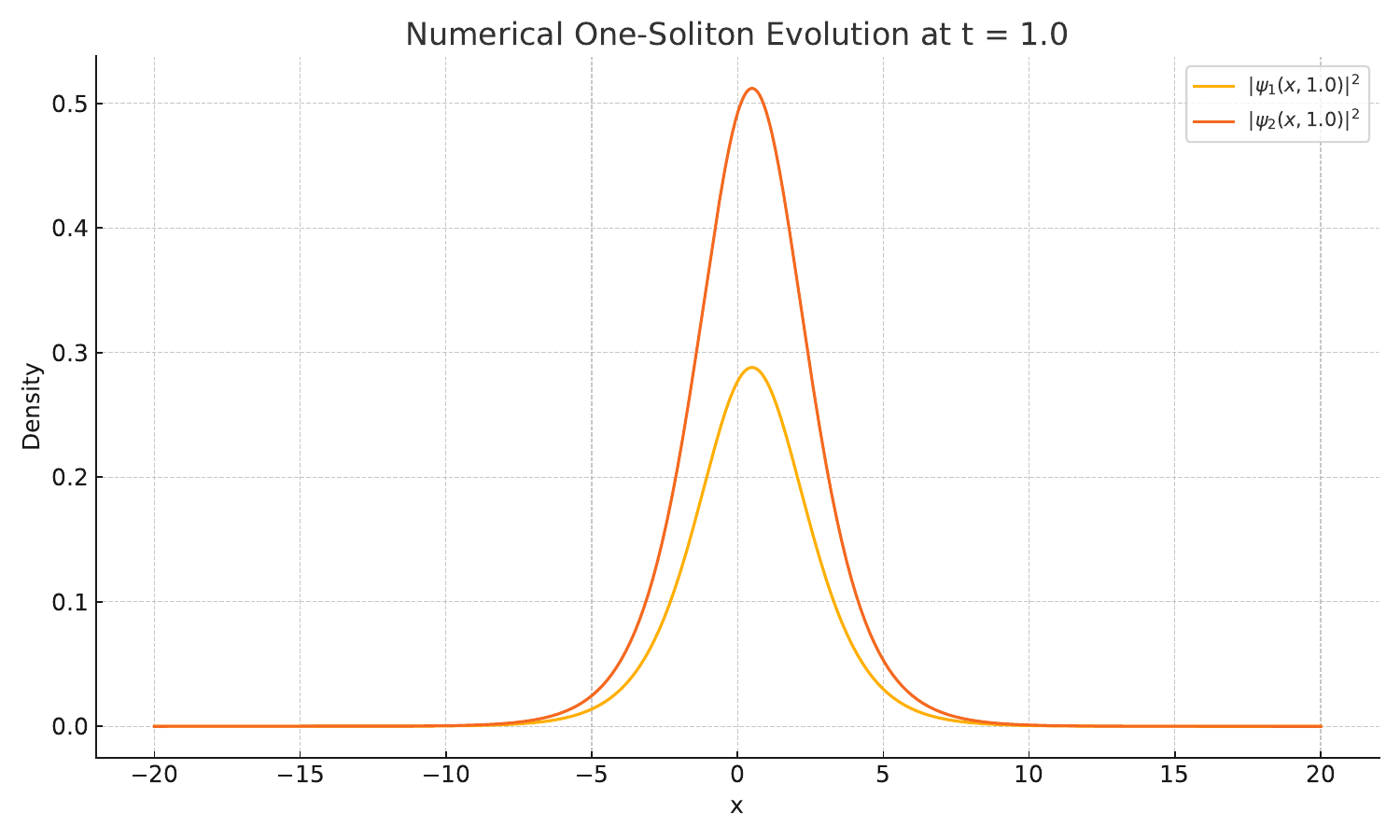} %
\includegraphics[width=0.65\linewidth]{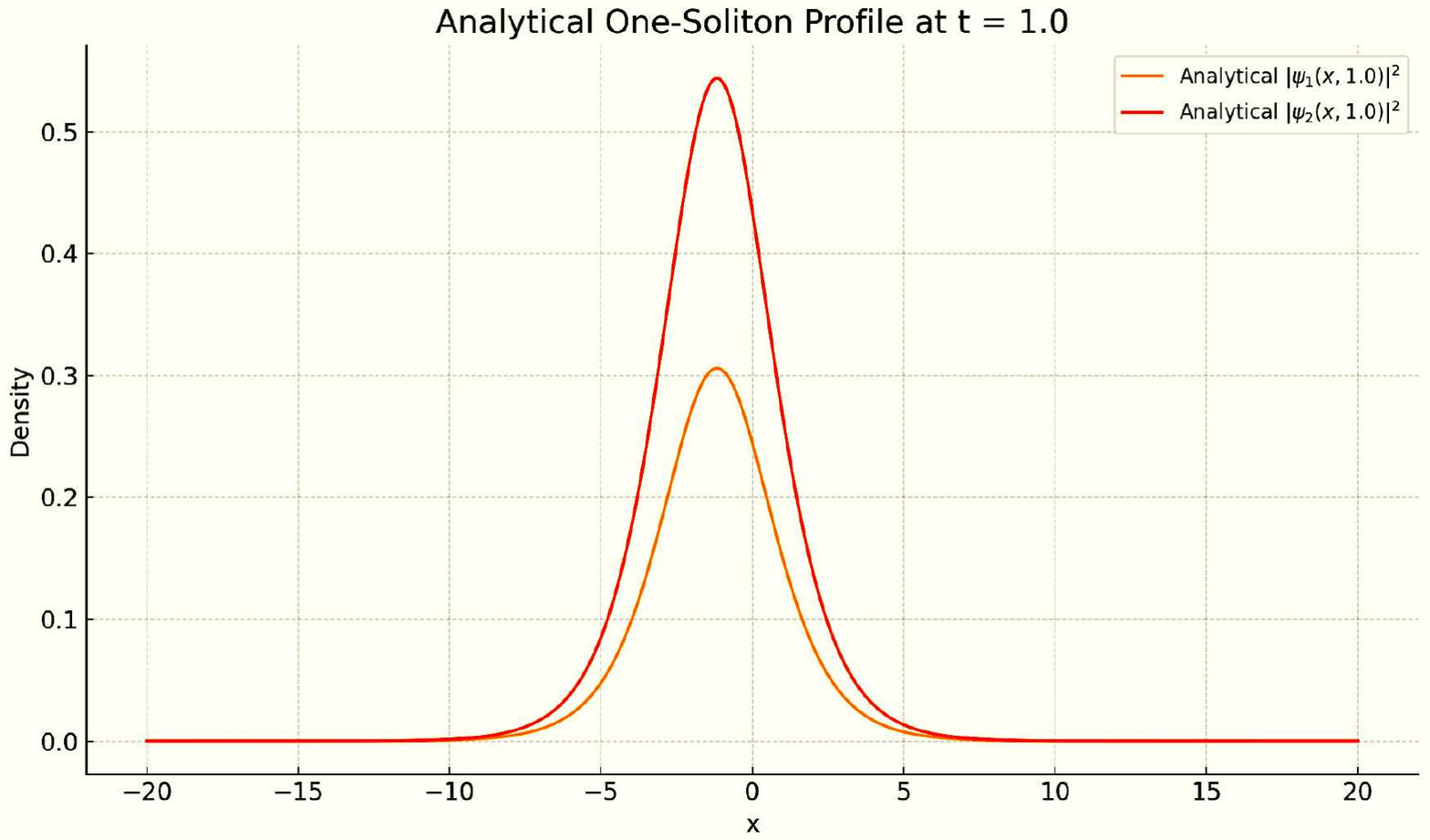}
\end{center}
\caption{ \textbf{The top panel}: The numerically evolved one-soliton state at $t =
1.0$ for both components $|\protect\psi_1(x,t)|^2$ and $|\protect\psi%
_2(x,t)|^2$. \textbf{The bottom panel}: The corresponding analytical one-soliton
solution evaluated at $t = 1.0$. The close agreement between the analytical
and numerical results confirms the accuracy of the analytical solution and
the stability of the system under time-dependent coefficients. The
parameters used are $\protect\varepsilon_1^{(1)}=0.6$, $\protect\varepsilon%
_2^{(1)}=0.8$ such that $|\protect\varepsilon_1^{(1)}|^2+|\protect\varepsilon%
_2^{(1)}|^2 =1$, $\protect\zeta_{10} = 0.4$, $\protect\zeta_{20} = 0.2$, $%
\protect\delta_1 = 0.1$, and $\protect\chi_1 = 0.2$. The coupling
coefficients are time-dependent: $\protect\alpha(t) = \protect\beta(t) =
0.1\cos(0.02t)$, and the phase modulation is given by $\Gamma(t) = -0.03t$.}
\end{figure}

\subsection{The two-soliton solution}

The gauge-transformation method can be easily extended to generate
multisoliton solutions. In particular, the two-component two-soliton
solution $\psi _{1,2}^{(2)}$ can be expressed as

\begin{eqnarray}
\psi _{1}^{(2)} &=&\psi _{1}^{(1)}-2i(\zeta _{2}-\bar{\zeta _{2}})\tilde{P}%
_{12}\frac{1}{\sqrt{\alpha (t)}}\mathrm{exp\left(i\Gamma (t)x^{2}/2\right)}
\label{ch6eq43} \\
\psi _{2}^{(2)} &=&\psi _{2}^{(1)}-2i(\zeta _{2}-\bar{\zeta _{2}})\tilde{P}%
_{13}\frac{1}{\sqrt{\beta (t)}}\mathrm{exp\left(i\Gamma (t)x^{2}/2\right)}
\label{ch6eq44}
\end{eqnarray}%
where $\zeta _{2}=\bar{\zeta _{2}}^{\ast }=\zeta _{21}+i\zeta _{22}$, with $%
\zeta _{21}(t)=\zeta _{210}\exp (-2\int \Gamma (t)\,dt)$, $\zeta
_{22}(t)=\zeta _{220}\exp (-2\int \Gamma (t)\,dt)$ while $\tilde{P}_{12}$
and $\tilde{P}_{13}$ are elements of the projection matrix derived from
matrix
\begin{equation}
M^{(2)}=\Phi ^{(0)}(x,t,\bar{\zeta}_{2})\cdot \left(
\begin{array}{ccc}
e^{2\delta _{2}}\sqrt{2} & \varepsilon _{1}^{(2)}e^{2i\chi _{2}} &
\varepsilon _{2}^{(2)}e^{2i\chi _{2}} \\
\varepsilon _{1}^{\ast (2)}e^{-2i\chi _{2}} & e^{-2\delta _{2}}/\sqrt{2} & 0
\\
\varepsilon _{2}^{\ast (2)}e^{-2i\chi _{2}} & 0 & e^{-2\delta _{2}}/\sqrt{2}
\\
&  &
\end{array}%
\right) \cdot \Phi ^{(0)}(x,t,\zeta _{2})^{-1}.
\end{equation}

Thus, the explicit form of the two-soliton solution can be written as
\begin{equation}
\psi_1^{(2)} =\frac{2\sqrt{2}}{\sqrt{\alpha(t)}} e^{-2 \int \Gamma(t) dt +
\frac{1}{2} i x^2 \Gamma(t)} \left( - \frac{e^{2\delta_1 + 2i\chi_1}
\varepsilon_{1}^{(1)} \zeta_{20}}{D_1} - \frac{e^{2\delta_2 + 2i\chi_2}
\varepsilon_{2}^{(1)} \zeta_{220}}{D_2} \right)
\end{equation}

\begin{equation}
\psi_2^{(2)} = \frac{2\sqrt{2}}{\sqrt{\beta(t)}} e^{-2 \int \Gamma(t) dt +
\frac{1}{2} i x^2 \Gamma(t)} \left( - \frac{e^{2\delta_1 +
2i\chi_1}\varepsilon_{2}^{(1)} \zeta_{20}}{D_1} - \frac{e^{2\delta_2 +
2i\chi_2} \varepsilon_{2}^{(2)} \zeta_{220}}{D_2} \right)
\end{equation}

where the denominators are

\begin{equation*}
D_{1}=\exp \left( 4\delta _{1}+4i\left[ \frac{1}{2}e^{-2\int \Gamma
(t)dt}x+\int e^{-4\int \Gamma (t)dt}dt(\zeta _{10}+i\zeta _{20})\right]
\left( \zeta _{10}+i\zeta _{20}\right) \right) +e^{4i\theta _{1}},
\end{equation*}

\begin{equation*}
D_{2}=\exp \left( 4\delta _{2}+4i\left[ \frac{1}{2}e^{-2\int \Gamma
(t)dt}x+\int e^{-4\int \Gamma (t)dt}dt(\zeta _{210}+i\zeta _{220})\right]
\left( \zeta _{210}+i\zeta _{220}\right) \right) +e^{4i\theta _{2}}.
\end{equation*}%
and
\begin{equation*}
\theta _{1}=\int e^{-4\int \Gamma (t)dt}dt\left( \zeta _{10}-i\zeta
_{20}\right) ^{2}+2e^{-2\int \Gamma (t)dt}x\left( i\zeta _{10}+\zeta
_{20}\right)
\end{equation*}

\begin{equation*}
\theta _{2}=\int e^{-4\int \Gamma (t)dt}dt\left( \zeta _{210}-i\zeta
_{220}\right) ^{2}+2e^{-2\int \Gamma (t)dt}x\left( i\zeta _{210}+\zeta
_{220}\right) .
\end{equation*}

\subsection{The three-soliton solution}

The three-soliton solution $\psi_{1,2}^{(3)}$ is constructed by applying the
gauge transformation a third time on the two-soliton solution $%
\psi_{1,2}^{(2)}$, resulting in:
\begin{eqnarray}
\psi_1^{(3)} &=& \psi_1^{(2)} - 2i(\zeta_3 - \bar{\zeta}_3)\, \hat{P}_{12}
\, \frac{1}{\sqrt{\alpha(t)}} \exp\left(i \Gamma(t) \frac{x^2}{2} \right),
\label{ch6eq45} \\
\psi_2^{(3)} &=& \psi_2^{(2)} - 2i(\zeta_3 - \bar{\zeta}_3)\, \hat{P}_{13}
\, \frac{1}{\sqrt{\beta(t)}} \exp\left(i \Gamma(t) \frac{x^2}{2} \right),
\label{ch6eq46}
\end{eqnarray}
where $\zeta_3 = \bar{\zeta}_3^* = \zeta_{31} + i \zeta_{32}$, and
\begin{equation*}
\zeta_{31}(t) = \zeta_{310} \exp\left(-2 \int \Gamma(t)\, dt\right), \quad
\zeta_{32}(t) = \zeta_{320} \exp\left(-2 \int \Gamma(t)\, dt\right).
\end{equation*}
The projection matrix elements $\hat{P}_{12}$ and $\hat{P}_{13}$ are derived
from the matrix $M^{(3)}$ as:
\begin{equation*}
M^{(3)} = \Phi^{(0)}(x,t,\bar{\zeta}_3) \cdot
\begin{pmatrix}
e^{2\delta_3} \sqrt{2} & \varepsilon_1^{(3)} e^{2i\chi_3} &
\varepsilon_2^{(3)} e^{2i\chi_3} \\
\varepsilon_1^{*(3)} e^{-2i\chi_3} & \dfrac{e^{-2\delta_3}}{\sqrt{2}} & 0 \\
\varepsilon_2^{*(3)} e^{-2i\chi_3} & 0 & \dfrac{e^{-2\delta_3}}{\sqrt{2}}%
\end{pmatrix}
\cdot \Phi^{(0)}(x,t,\zeta_3)^{-1}.
\end{equation*}

Thus, the three-soliton state takes the following general form:

\begin{equation}
\psi^{(3)}_{1,2}(x,t) = \frac{2 \sqrt{2} e^{-2 I + \frac{1}{2} i x^2 \Gamma
(t)}}{\sqrt{\pm \alpha (t)}} \sum_{j=1}^{3} -\frac{\zeta_{j0}
\varepsilon_{j1} e^{2 \delta_j + 2 i \chi_j}}{D_{3j}(x,t)}
\end{equation}
and the denominator $D_{3j}(x,t) $ is

\begin{eqnarray}
D_{3j}(x,t) = \exp\left(4 \delta_j + 4 i (\zeta_{j0} + i \zeta_{j20}) \left(%
\frac{1}{2} x e^{-2 I} + (\zeta_{j0} + i \zeta_{j20}) J \right)\right)
\notag \\
+ \exp\left(4 i (\zeta_{j0} - i \zeta_{j20})^2 J + 2 (\zeta_{j20} + i
\zeta_{j0}) x e^{-2 I} \right)  \notag
\end{eqnarray}
where $J = \int e^{-4 I} dt$, $I = \int \Gamma (t) dt$. Here, the $\pm $ in
the denominator corresponds to $+\alpha (t) $ for $\psi_1^{(3)}$ and $%
-\alpha (t)=\beta(t)$ for $\psi_2^{(3)}$.

\subsection{The four-soliton solution}

By applying the gauge transformation a fourth time, the four-soliton
solution $\psi_{1,2}^{(4)}$ is obtained:
\begin{eqnarray}
\psi_1^{(4)} &=& \psi_1^{(3)} - 2i(\zeta_4 - \bar{\zeta}_4)\, \bar{P}_{12}
\, \frac{1}{\sqrt{\alpha(t)}} \exp\left(i \Gamma(t) \frac{x^2}{2} \right),
\label{ch6eq47} \\
\psi_2^{(4)} &=& \psi_2^{(3)} - 2i(\zeta_4 - \bar{\zeta}_4)\, \bar{P}_{13}
\, \frac{1}{\sqrt{\beta(t)}} \exp\left(i \Gamma(t) \frac{x^2}{2} \right),
\label{ch6eq48}
\end{eqnarray}
where $\zeta_4 = \bar{\zeta}_4^* = \zeta_{41} + i \zeta_{42}$, and
\begin{equation*}
\zeta_{41}(t) = \zeta_{410} \exp\left(-2 \int \Gamma(t)\, dt\right), \quad
\zeta_{42}(t) = \zeta_{420} \exp\left(-2 \int \Gamma(t)\, dt\right).
\end{equation*}
The projection matrix for the fourth soliton is derived from the matrix $%
M^{(4)}$ as:
\begin{equation*}
M^{(4)} = \Phi^{(0)}(x,t,\bar{\zeta}_4) \cdot
\begin{pmatrix}
e^{2\delta_4} \sqrt{2} & \varepsilon_1^{(4)} e^{2i\chi_4} &
\varepsilon_2^{(4)} e^{2i\chi_4} \\
\varepsilon_1^{*(4)} e^{-2i\chi_4} & \dfrac{e^{-2\delta_4}}{\sqrt{2}} & 0 \\
\varepsilon_2^{*(4)} e^{-2i\chi_4} & 0 & \dfrac{e^{-2\delta_4}}{\sqrt{2}}%
\end{pmatrix}
\cdot \Phi^{(0)}(x,t,\zeta_4)^{-1}.
\end{equation*}

The explicit form of the four-soliton solution is

\begin{equation}
\psi^{(4)}_{1,2}(x,t) = \frac{2 \sqrt{2} e^{-2 I + \frac{1}{2} i x^2 \Gamma
(t)}}{\sqrt{\pm \alpha (t)}} \sum_{j=1}^{4} -\frac{\zeta_{j0}
\varepsilon_{j1} e^{2 \delta_j + 2 i \chi_j}}{D_{4j}(x,t)}
\end{equation}
where $I = \int \Gamma (t) dt$, $J = \int e^{-4 I} dt$, and the denominator $%
D_{4j}(x,t) $ is

\begin{eqnarray}
D_{4j}(x,t) = \exp\left( 4 \delta_j + 2 i (\zeta_{j0} + i \zeta_{j20})
e^{-2I} \left( x + 2 (\zeta_{j0} + i \zeta_{j20}) e^{2I} J \right) \right) \\
+ \exp\left( 2 (\zeta_{j0} - i \zeta_{j20}) e^{-2I} \left( 2 (\zeta_{j20} +
i \zeta_{j0}) e^{2I} J + i x \right) \right).  \notag
\end{eqnarray}

Note that the constraint $|\varepsilon^{(j)}_1|^2 + |\varepsilon^{(j)}_2|^2
= 1$ for $j = 1, 2, 3, 4$ applies to the coupling parameters used in
constructing the multi-soliton solution. These parameters determine the
internal distribution of amplitude between the two components in each
soliton. This relation does not impose a direct constraint on the physical
condensate densities $|\psi_1(x,t)|^2$ and $|\psi_2(x,t)|^2$, which evolve
according to the full analytical solution.

\section{Dynamics of the four-soliton solution}

\subsection{General consideration}

The solution given in Eq.(39) demonstrates that the amplitude of $\psi
_{1,2}^{(4)}(x,t)$ is mainly affected by the exponential prefactor $e^{-2I}$%
, which depends on $\Gamma (t)$ via $I=\int \Gamma (t)dt$. This term can
introduce growth, decay, or oscillations. The denominator $D_{4j}(x,t)$
governs the location and interactions of the soliton. If $\Gamma (t)$ varies
in time, the overall amplitude of $\psi _{1,2}^{(4)}(x,t)$ changes too. If $%
\Gamma (t)$ is constant, the soliton's amplitudes may only change due to
interactions.

The peak location of the solitons can be determined as $x_{j}^{\mathrm{peak}%
}(t)\approx -2(\zeta _{j0}+i\zeta _{j20})e^{2I}J$. Because $J=\int e^{-4I}dt$%
, the position of the soliton's maximum depends on the integral of $e^{-4I}$%
. The velocity is obtained as $v_{j}(t)=dx_{j}^{\mathrm{peak}}/dt$. The
time-dependent factor $e^{2I}$ means that the soliton may accelerate or
decelerate, depending on $\Gamma (t)$. The soliton's phase is $\phi
(x,t)=-2I+\frac{1}{2}x^{2}\Gamma (t)+\sum_{j=1}^{4}(2\chi _{j}-\arg
(D_{4j}(x,t))$. The first term, $-2I$, evolves with $\Gamma (t),$
introducing a global phase shift. The quadratic term $\frac{1}{2}x^{2}\Gamma
(t)$ suggests that $\Gamma (t)$ contributes to the quadratic chirp,
modifying the wavefront curvature. The argument of $D_{4j}(x,t)$ affects
phase shifts produced by interactions between solitons.

We can now suitably tune the trapping frequency $\lambda ^{2}(t)=\Gamma
^{2}(t)+\Gamma ^{\prime }(t)/2$ and the intra-/inter-component interaction
coefficients $\alpha (t)$, to reveal the nonlinear excitations. The
polarization coefficients, such as $\varepsilon _{1}^{(1)}$ and its
counterparts, determine the population balance between soliton states.
\textbf{These coefficients can be interpreted as probability amplitudes in
the two-component BEC, which represent quantum states in the qubit basis
\cite{tim2012}. Specifically, the state of the soliton can be written as $%
|\psi \rangle =\varepsilon _{1}^{(1)}|1\rangle +\varepsilon
_{2}^{(1)}|0\rangle $, demonstrating a direct similarity to the qubit
representation. By carefully tuning the polarization parameters, it is
possible to manipulate the quantum state of the system. Hence, this
framework offers a classical counterpart of the qubit-like behavior, which
may be useful for simulating the basic quantum logic} \cite{tim2012,
kony2025, shaukat2017, tvngo2021}. The parameters $\zeta _{i0}$, $\delta
_{i} $, and $\Gamma (t)$ play a crucial role in determining how solitons
separate and interact over time. The expression for the velocity, $%
v_{i}=-2\zeta _{i0}\exp \left( -2\int \Gamma (t)dt\right) $, suggests that
adjusting $\zeta _{i0}$ or the time-dependent function $\Gamma (t)$ can
effectively control the speed at which solitons move apart, affecting the
fission rate. Meanwhile, amplitude $A_{i}=2\sqrt{2}\zeta _{i0}e^{2\delta
_{i}}$ determines the energy distribution between the emerging solitons,
meaning that varying $\delta _{i}$ can control post-fission intensity
variations. By fine-tuning these parameters, it is possible to modulate
soliton interactions, delay or accelerate the fission, and even regulate
energy transfer between solitonic components.

\subsection{Superposition states of four solitons}

Figure~2 illustrates the superposition dynamics of the four-soliton state in
a two-component BEC. The top panel shows the individual density
distributions of both components $|\psi _{1}(x,t)|^{2}$ and $|\psi
_{2}(x,t)|^{2}$ of Eq.(39), where the interference patterns and energy
sharing between solitons in each component are evident. The bottom panel
presents a contour plot of the superposition state visualizing the evolving
structure of the four-soliton interaction. The periodic modulation of the
interaction strength, $\alpha (t)=0.1\cos (0.02t)$, stabilizes the soliton
dynamics, while the linear phase shift function, $\Gamma (t)=-0.03t$,
contributes to global temporal phase modulation. The real parts of the
complex spectral parameters, which determine the soliton velocities, are
chosen as $\zeta _{10}=0.2$, $\zeta _{20}=0.01$, $\zeta _{210}=0.4$, $\zeta
_{220}=0.04$, $\zeta _{310}=0.6$, $\zeta _{320}=0.06$, $\zeta _{410}=0.8$,
and $\zeta _{420}=0.08$. These values set distinct group velocities and
allow for an observable superposition during propagation. The amplitude
scaling factors $\delta _{1}=0.1$, $\delta _{2}=0.4$, $\delta _{3}=0.6$, and
$\delta _{4}=0.08$ control the energy and peak height of each soliton. Phase
offsets between the components are tuned using $\chi _{1}=0.01$, $\chi
_{2}=0.03$, $\chi _{3}=0.05$, and $\chi _{4}=0.7$. During the central
interval, the solitons exhibit coherent overlap and interference, resembling
quantum superposition. Outside this regime, the solitons separate,
analogously modeling quantum decoherence where entangled states collapse
into distinguishable ones. This dynamics demonstrates the potential for
simulating quantum logic operations within classical nonlinear systems \cite%
{kanna2001, tim2012, kony2025, shaukat2017, tvngo2021}.

\begin{figure}[tbp]
\begin{center}
\includegraphics[width=0.475\linewidth]{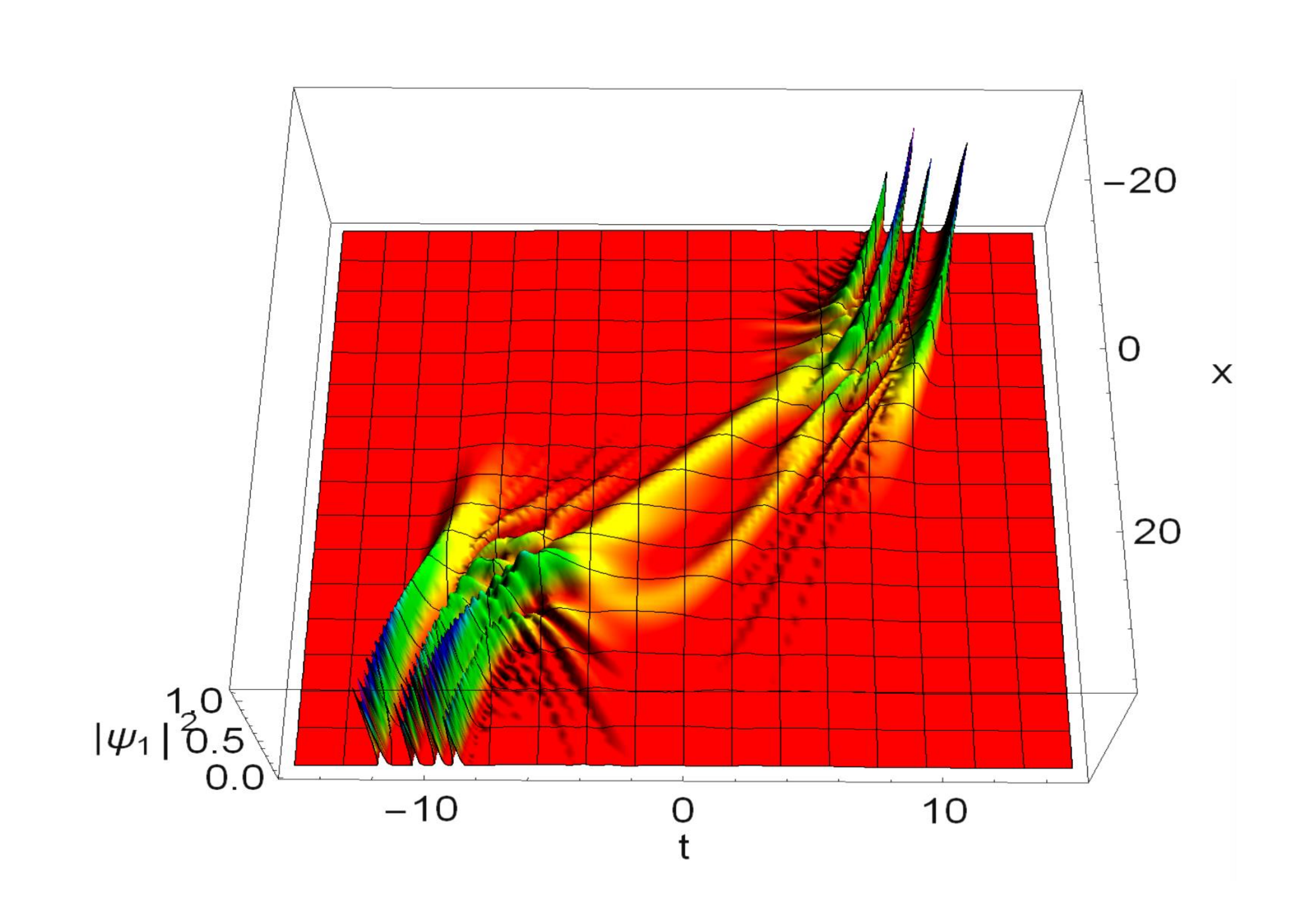} %
\includegraphics[width=0.475\linewidth]{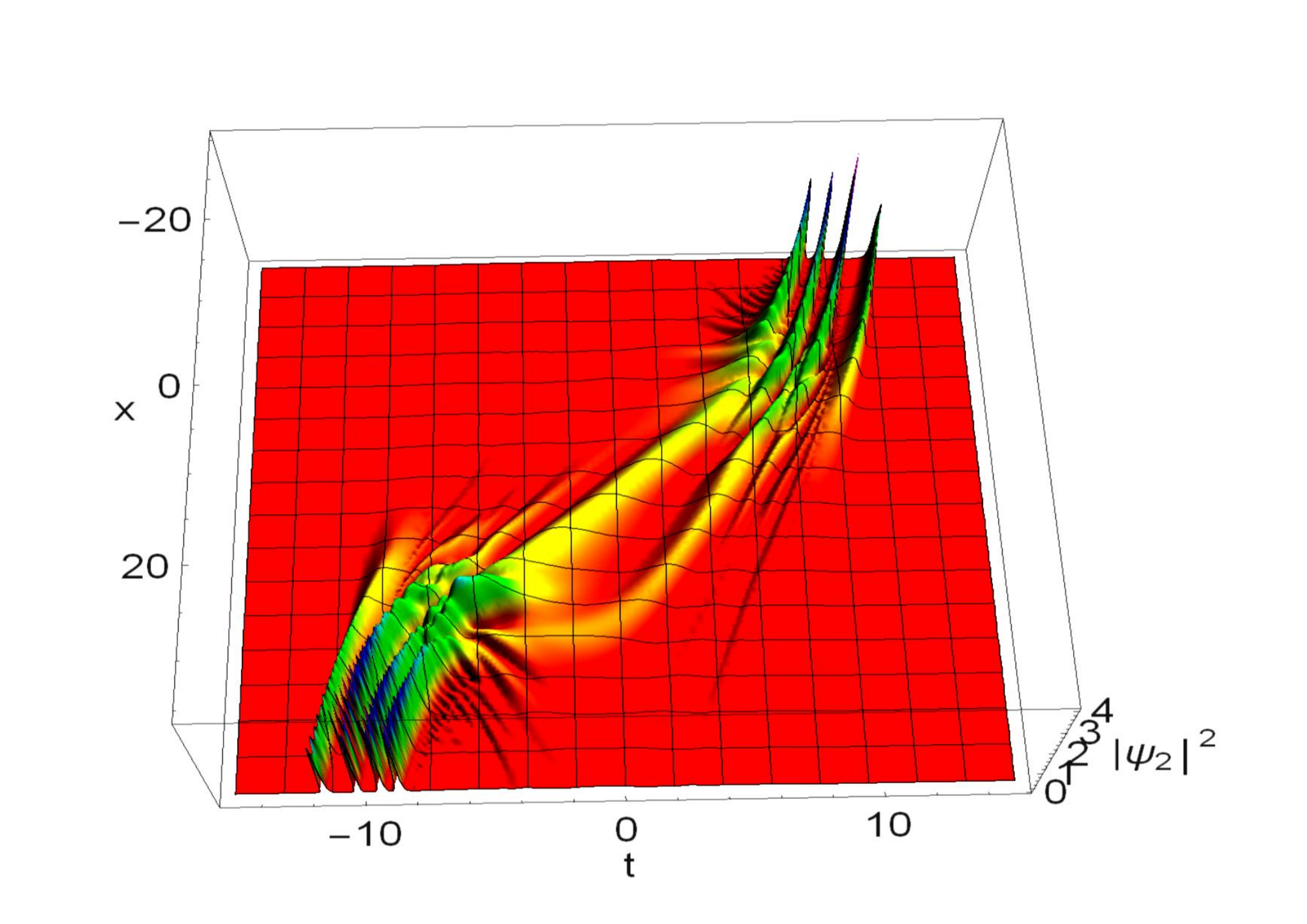} %
\includegraphics[width=0.45\linewidth]{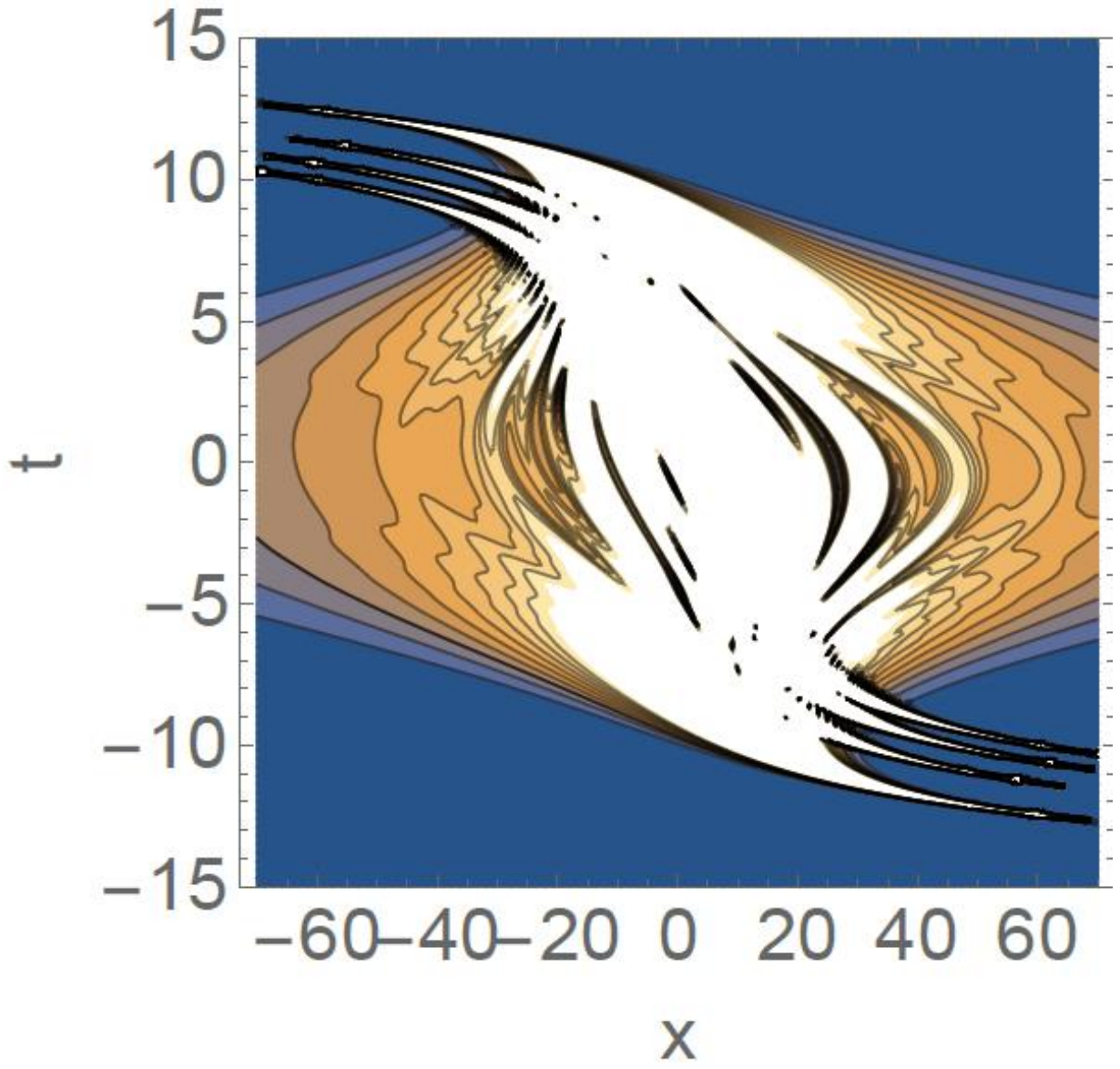} %
\includegraphics[width=0.5\linewidth]{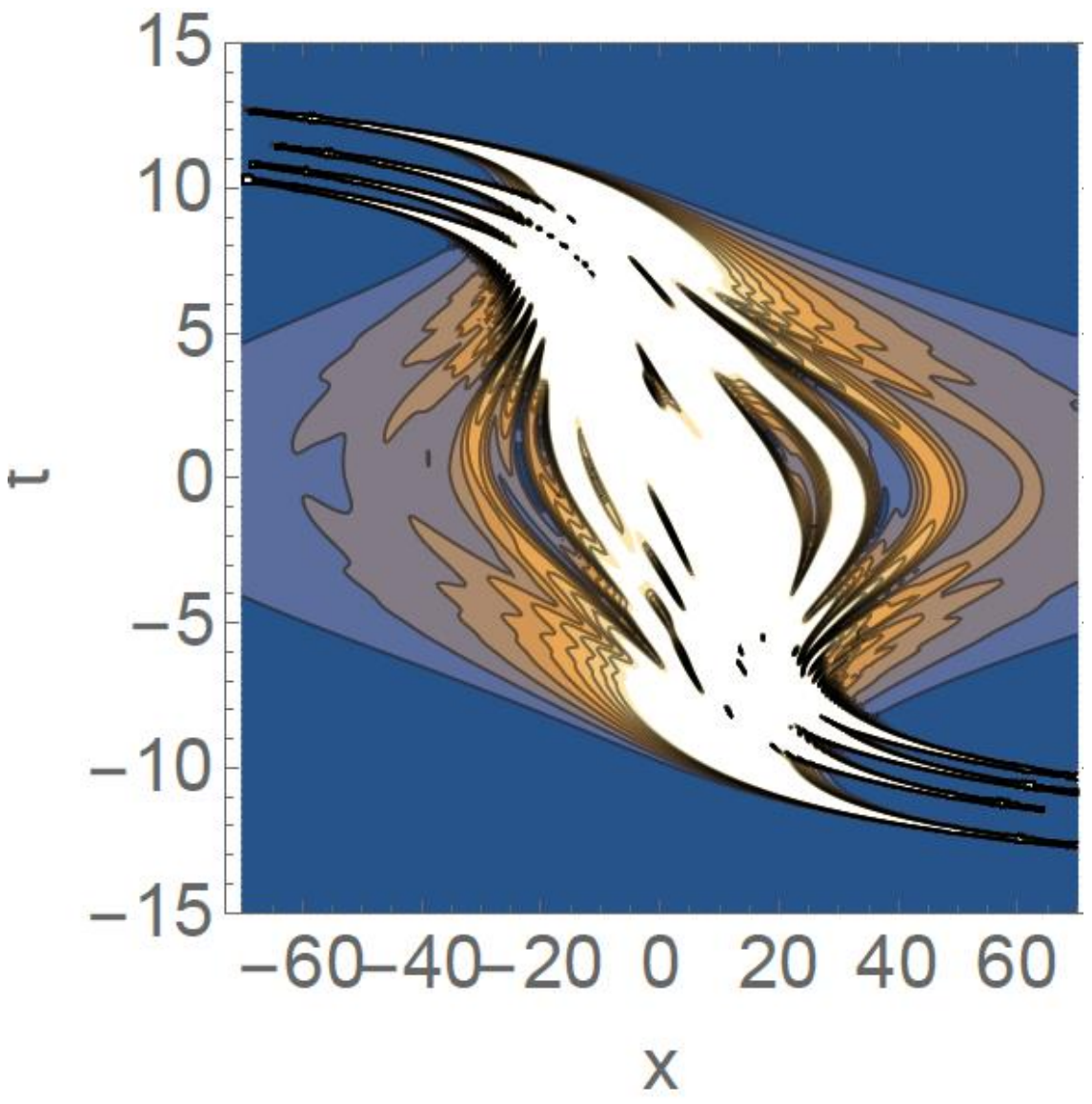}
\end{center}
\caption{\textbf{The top panel}: Density plots of the individual components $|\protect%
\psi_1(x,t)|^2$ and $|\protect\psi_2(x,t)|^2$ given by eqn. (39), showing
the evolution and superposition state of four-soliton in a two-component
Bose--Einstein condensate. \textbf{The bottom panel}: The contour plot of the
superposition state highlighting the coherent soliton interference. The
interaction strength is modulated by $\protect\alpha(t) = 0.1 \cos(0.02t)$,
and the phase function is governed by $\Gamma(t) = -0.03t$. Parameters used:
$\protect\delta_1 = 0.1 $, $\protect\delta_2 = 0.4$, $\protect\delta_3 = 0.6$%
, $\protect\delta_4 = 0.08$; $\protect\zeta_{10} = 0.2$, $\protect\zeta_{20}
= 0.01$, $\protect\zeta_{210} = 0.4$, $\protect\zeta_{220} = 0.04$, $\protect%
\zeta_{310} = 0.6$, $\protect\zeta_{320} = 0.06$, $\protect\zeta_{410} = 0.8$%
, $\protect\zeta_{420} = 0.08$; $\protect\chi_1 = 0.01$, $\protect\chi_2 =
0.03$, $\protect\chi_3 = 0.05$, $\protect\chi_4 = 0.7$; $\protect\varepsilon%
_{1}^{(1)} = 0.4$, $\protect\varepsilon_{1}^{(2)} = 0.2$, $\protect%
\varepsilon_{1}^{(3)} = 0.3 $, $\protect\varepsilon_{1}^{(4)} = 0.2$.}
\label{fig:superposition}
\end{figure}

\begin{figure}
\begin{center}
\includegraphics[width=0.55\linewidth]{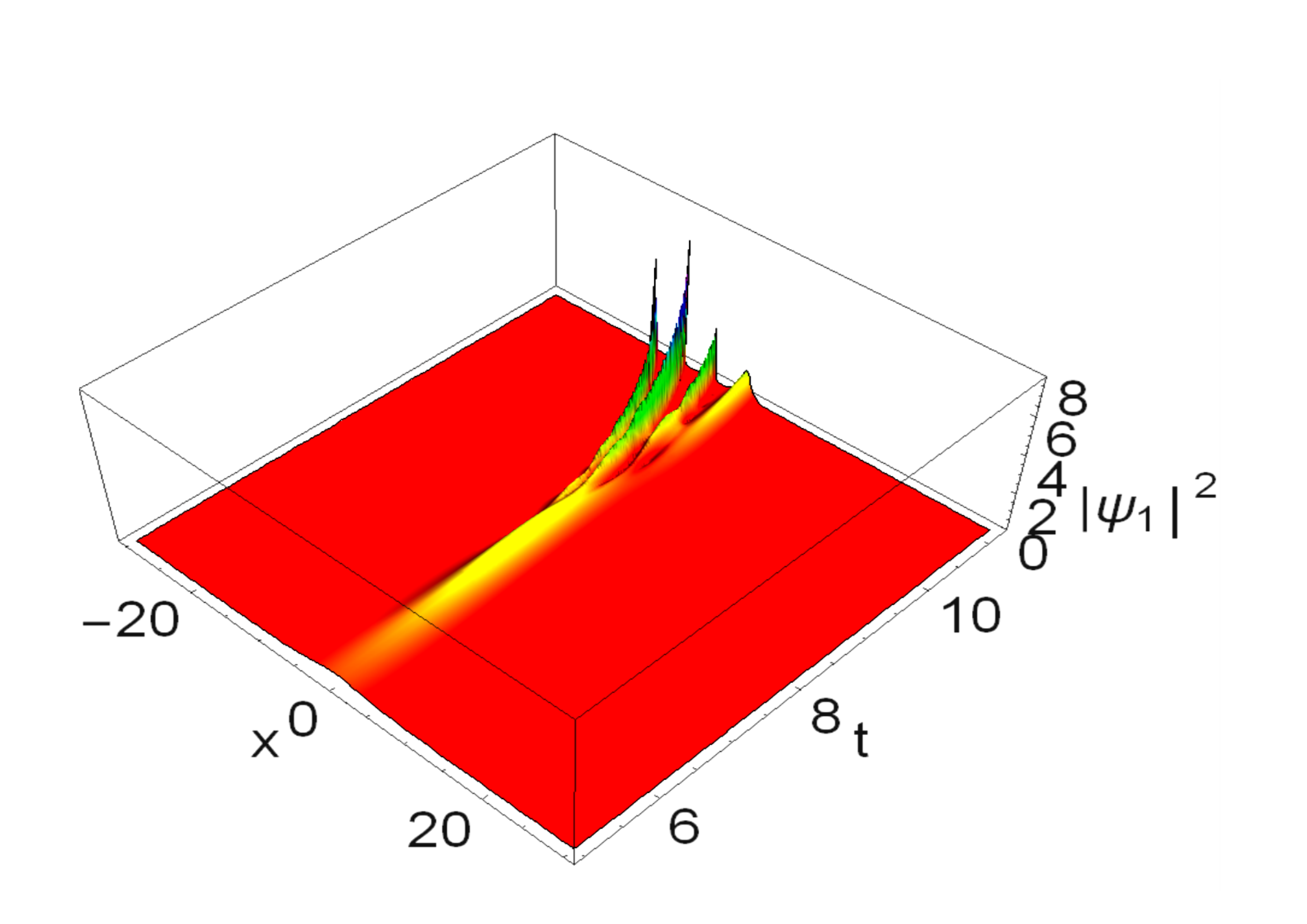}
\includegraphics[width=0.4\linewidth]{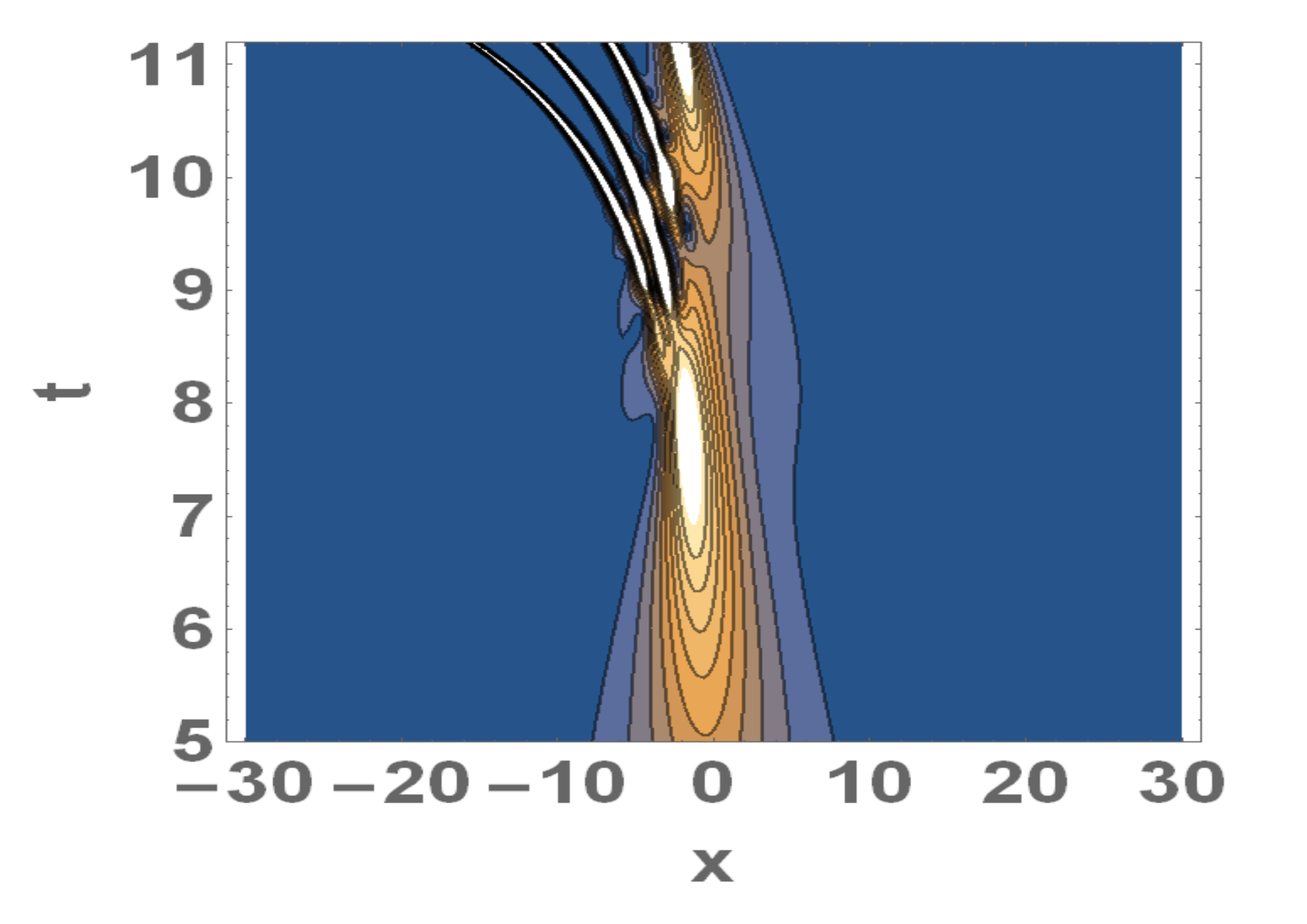}
\includegraphics[width=0.55\linewidth]{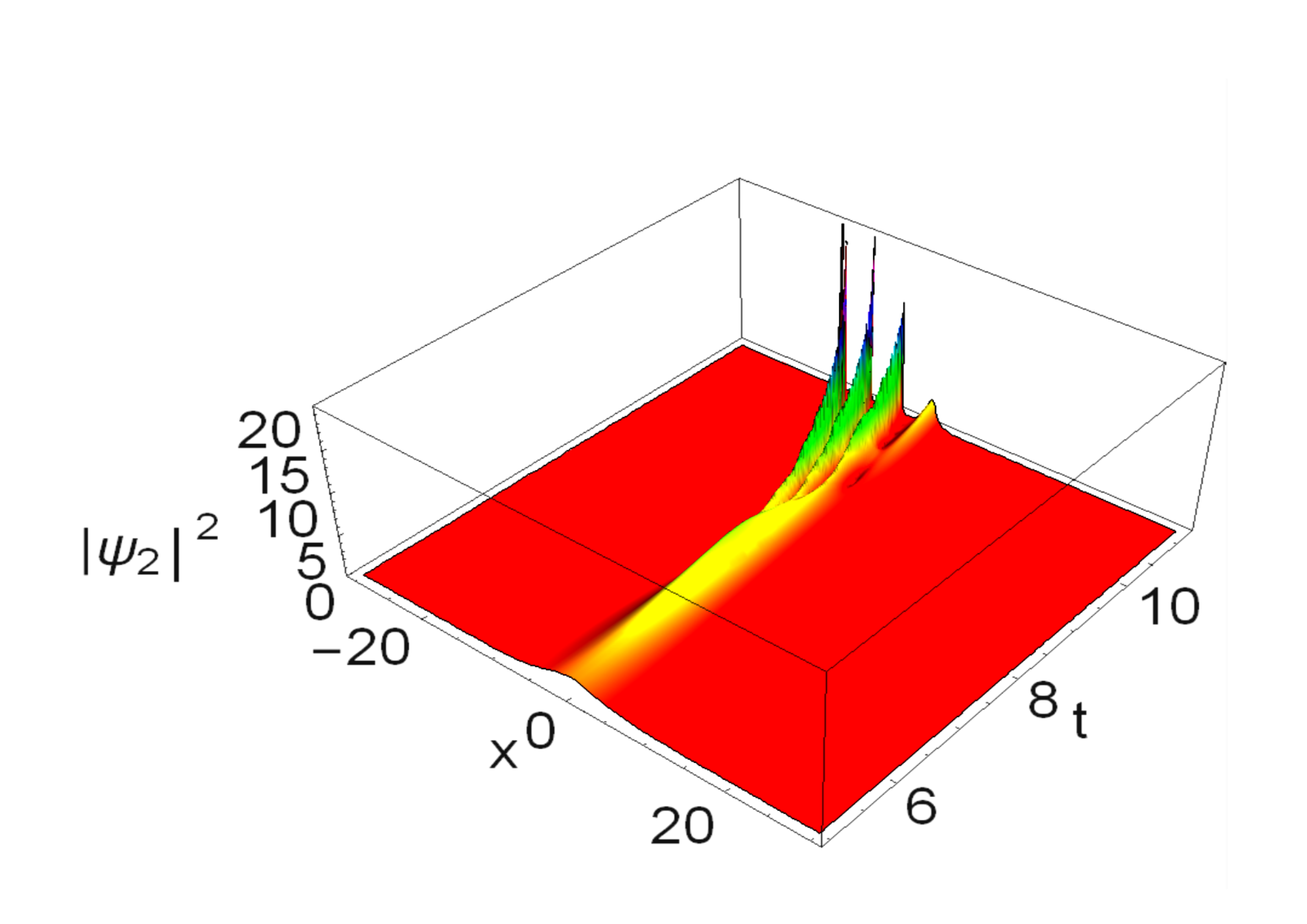}
\includegraphics[width=0.4\linewidth]{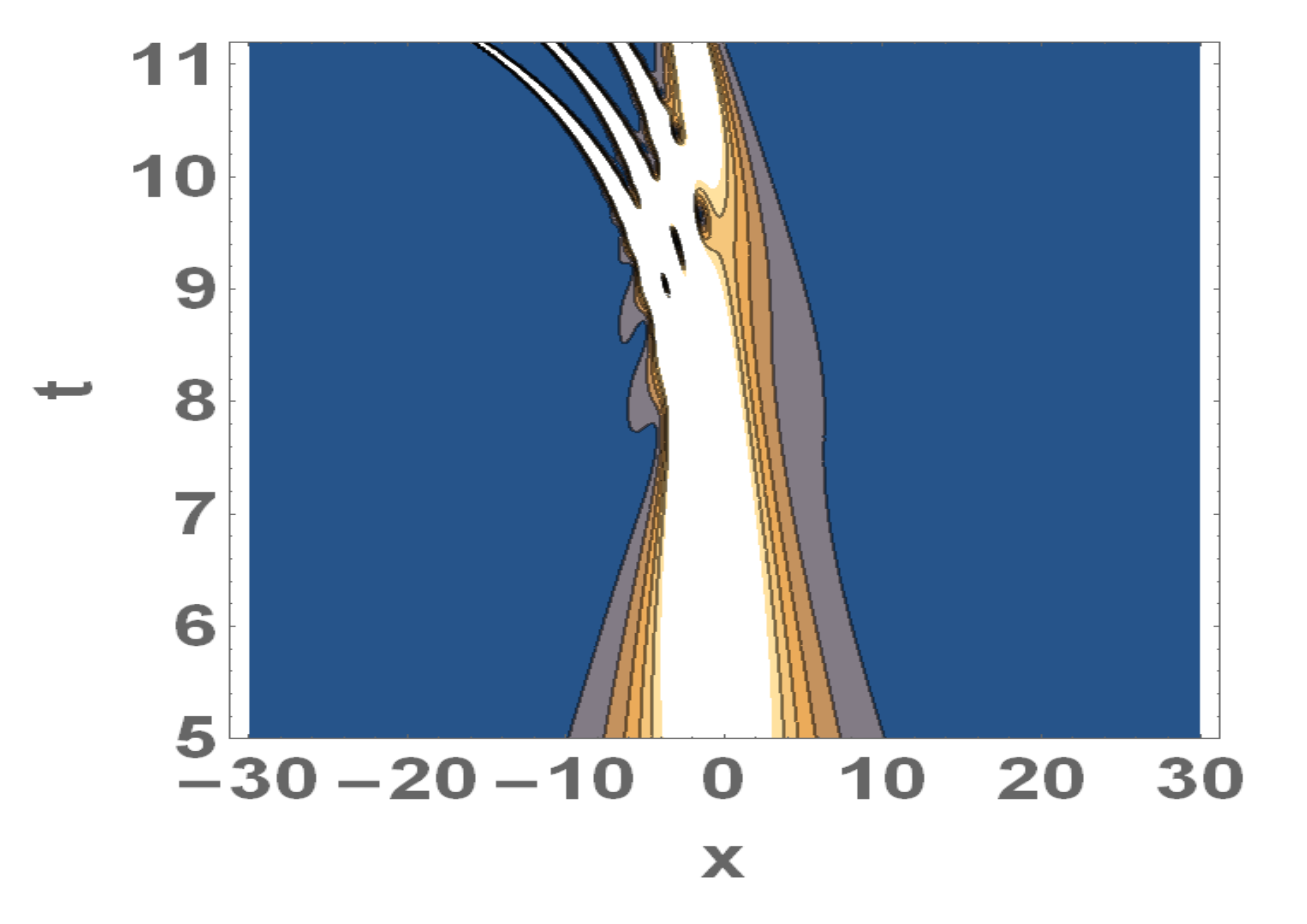}

\caption{The density plot (left) and the corresponding contour plot (right) given by eqn.(39) of the controlled four-soliton
fission for parameters $\delta _1=0.01$, $\delta _2=0.04$, $\delta _3=0.06$, $\delta _4=0.08$,
$\zeta _{10}=0.01$, $\zeta _{20}=0.01$, $\zeta _{210}=0.03$, $\zeta _{220}=0.04$, $\zeta _{310}=0.05$,
$\zeta _{320}=0.06$, $\zeta _{410}=0.07$, $\zeta _{420}=0.08$, $\chi _1=0.01$, $\chi _2=0.03$, $\chi _3=0.05$,
$\chi _4=0.07$, $\varepsilon _{1}^{(1)}=0.4$, $\varepsilon _{1}^{(2)}=0.2$,  $\varepsilon _{1}^{(3)}=0.3$,
$\varepsilon _{1}^{(4)}=0.2$, $\Gamma (t)=-0.035 t$, $\alpha (t)=0.1 \cos (0.002 t).$}
\end{center}
\end{figure}

\subsection{Controlled fission of four-soliton states}

Figure~3 illustrates the controlled fission of a four-soliton state of
two-component BEC. The density and contour plots show that the solitons are
initially bound and gradually separate over time due to the influence of
time-dependent system parameters. The intra- and inter-species interaction
coefficients, modulated by $\alpha (t)=0.1\cos (0.002t)$, regulate the
soliton interactions and ensure a well-controlled fission process. The
external potential, defined by the phase function $\Gamma (t)=-0.035t$,
introduces a linear time-dependent phase shift that alters the trajectories
of the soliton peaks. The real parts of the complex spectral parameters are
selected as $\zeta_{10} = 0.01$, $\zeta_{20} = 0.01$, $\zeta_{210} = 0.03$, $%
\zeta_{220} = 0.04$, $\zeta_{310} = 0.05$, $\zeta_{320} = 0.06$, $%
\zeta_{410} = 0.07$, and $\zeta_{420} = 0.08$, which define the soliton
velocities and result in slightly separated propagation paths. The amplitude
parameters $\delta_1 = 0.01$, $\delta_2 = 0.04$, $\delta_3 = 0.06$, and $%
\delta_4 = 0.08$ control the initial pulse strengths and energy content of
each soliton. Phase constants $\chi_1 = 0.01$, $\chi_2 = 0.03$, $\chi_3 =
0.05$, and $\chi_4 = 0.07$ regulate the initial internal phase shifts of
each component. The polarization coefficients $\varepsilon_1^{(1)} = 0.4$, $%
\varepsilon_1^{(2)} = 0.2$, $\varepsilon_1^{(3)} = 0.3$, and $%
\varepsilon_1^{(4)} = 0.2$ determine the population imbalance between the
two components of the system and enable distinct soliton identities. The
evolution of the peak positions, governed by the relation
\begin{equation*}
x_{j}(t)=-2\zeta _{j0}\exp \left( -2\int \Gamma (t)\,dt\right) J(t),
\end{equation*}%
shows that each soliton moves away from the origin gradually, indicating
controlled spatial separation. Throughout this process, the soliton
amplitudes remain approximately constant in time, demonstrating the
preservation of localized energy and coherence, even as the solitons
dynamically redistribute in space. This controllable fission mechanism may
find direct applications in quantum information processing, where soliton
states can be manipulated similar to qubits in quantum computing. Recent
studies have demonstrated the potential of soliton-based qubit states for
metrological applications, highlighting the versatility of the soliton
dynamics in quantum technologies \cite{tvngo2021}.\newline
Figure 4 presents a symmetric four-soliton fission process, where the
solitons separate evenly in both positive and negative $x $-directions.
Unlike the asymmetric propagation in Fig. 3, this case is achieved by
choosing symmetric initial soliton parameters (e.g., $\zeta_{10} = -0.01 $, $%
\zeta_{20} = 0.01 $, $\zeta_{30} = -0.05 $, $\zeta_{40} = 0.07 $). The
periodic modulation of the interaction strength, \textit{viz.}, $\alpha(t) =
0.1\cos(0.2t) $, introduces oscillatory behavior that affects the
soliton-separation dynamics. The phase function $\Gamma(t) = -0.035t $
further modulates the relative phase differences, impacting the symmetry of
the fission. This symmetric splitting is significant for applications to
nonlinear optics and matter-wave engineering, where the stability and
coherence are crucially important for the robust wave propagation.
Additionally, the symmetric fission provides a physical analogy to
controlled quantum-gate operations, where entangled soliton components
evolve into separate, distinguishable quantum states, highlighting the
potential of using the two-component solitons for quantum information
applications. Recent experimental and theoretical studies have revealed
soliton complexes composed of multiple solitons in close spatial proximity,
resulting in frequent and strong interactions within two-component BECs.
These dense configurations offer important insights into the dynamics of
multi-soliton collisions and collective behavior \cite{smossman}.

\subsection{Elastic collision of four-soliton states}

Figure 5 depicts the elastic collision of a four-soliton state in the
two-component BEC under the action of the time-dependent potential. The
solitons interact and emerge with their initial shape and velocity
preserved, confirming the integrability of the system. The soliton dynamics
are affected by the time-dependent scattering length $\alpha (t)=0.1e^{0.02t}
$, which gradually modulates the interaction strength, securing the stable
evolution. Additionally, the weak phase shift introduced by $\Gamma
(t)=-0.002t$ contributes to minor trajectory variations without disrupting
the coherence. The initial parameters are chosen to generate four distinct
soliton pulses: $\delta _{1}=0.75$, $\delta _{2}=0.82$, $\delta _{3}=0.69$,
and $\delta _{4}=0.85$ control the amplitude scaling of each soliton. The
velocity parameters are given by $\zeta _{10}=0.1$, $\zeta _{20}=0.2$, $%
\zeta _{210}=0.3$, $\zeta _{220}=0.4$, $\zeta _{310}=0.5$, $\zeta _{320}=0.6$%
, $\zeta _{410}=0.7$, and $\zeta _{420}=0.8$, which determine the
propagation speed of each soliton. The relative phase parameters are set as $%
\chi _{1}=0.9$, $\chi _{2}=0.75$, $\chi _{3}=0.8$, and $\chi _{4}=0.9$. The
polarization coefficients $\varepsilon _{1}^{(1)}=0.6$, $\varepsilon
_{1}^{(2)}=0.5$, $\varepsilon _{1}^{(3)}=0.4$, and $\varepsilon
_{1}^{(4)}=0.7$ determine the population distribution across the two
components. The external modulation is given by a weakly time-dependent trap
$\Gamma (t)=-0.002t$ and nonlinear coefficient $\alpha (t)=0.1\exp (0.02t)$.
The figure demonstrates that the solitons undergo elastic interactions, they
pass through each other without distortion, and their shapes and velocities
are preserved after collision. The elastic nature of the collision, in which
the solitons retain their identity post-interaction, contrasts with
inelastic processes that lead to fusion or decay. This soliton-collision
mode may have significant implications in nonlinear optics and quantum
information, where soliton-like wave packets can act as robust carriers of
quantum information, undergoing coherent transformations without loss of the
integrity.
\begin{figure}[tbp]
\begin{center}
\includegraphics[width=0.45\linewidth]{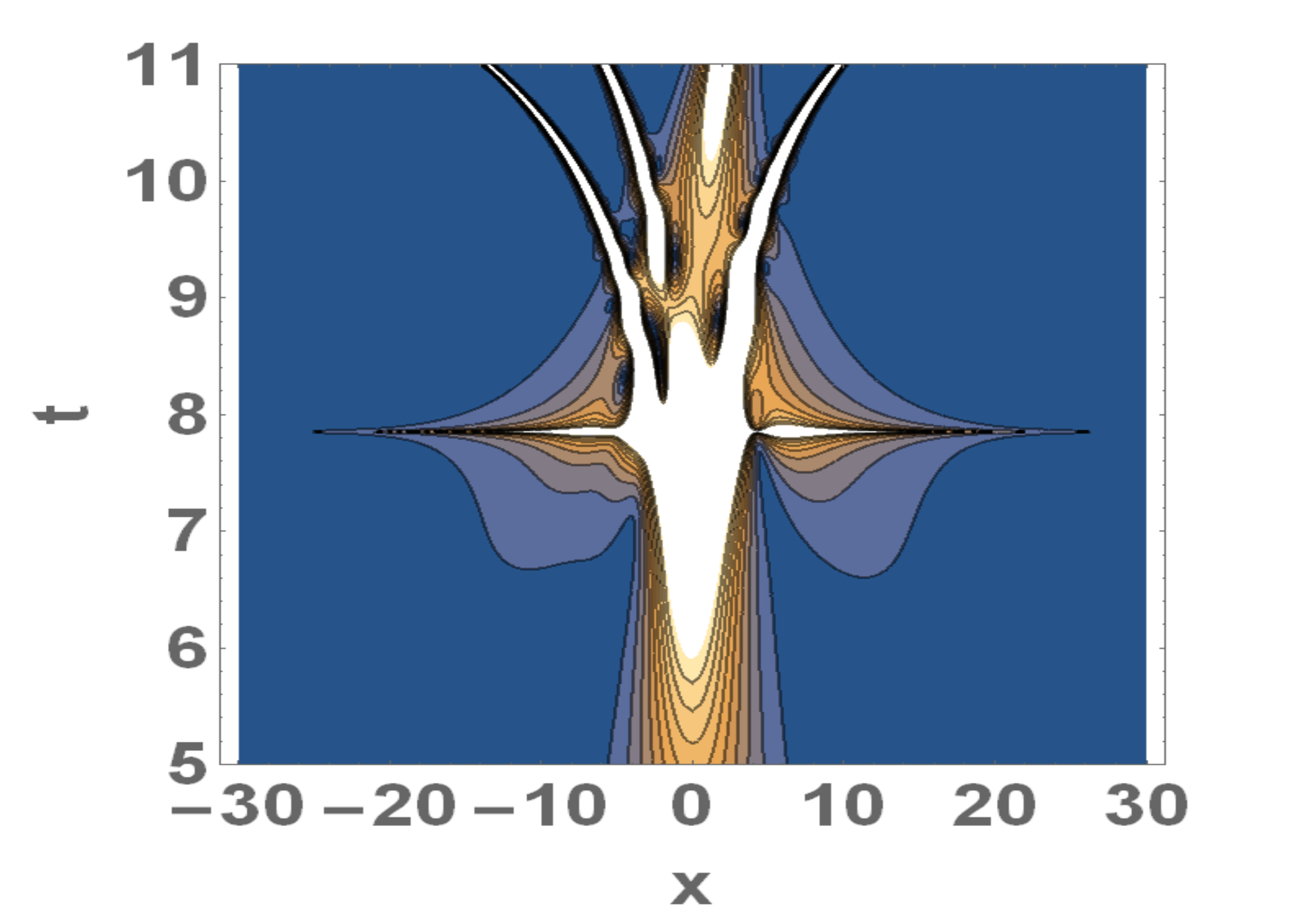} %
\includegraphics[width=0.45\linewidth]{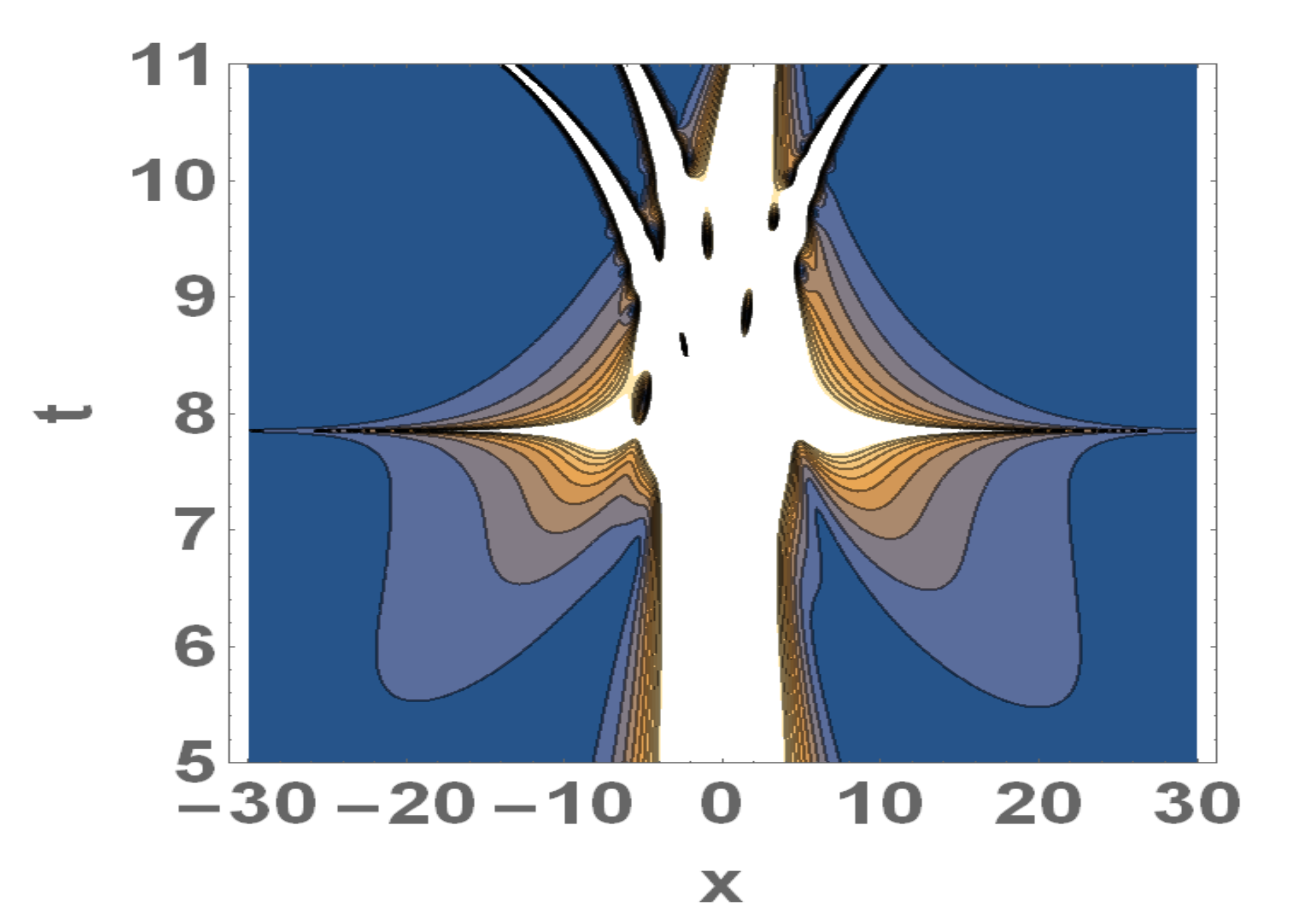}
\end{center}
\caption{Contour plots of the four-soliton symmetric fission dynamics in the
two-component Bose--Einstein condensate. \textbf{The left panel:} The
contour plot of component $|\protect\psi _{1}(x,t)|^{2}$. \textbf{The right
panel:} The contour plot of component $|\protect\psi _{2}(x,t)|^{2}$. The
plots demonstrate the symmetric separation and spatiotemporal evolution of
the soliton peaks under the temporal modulation. The parameters used are: $%
\protect\delta _{1}=0.01$, $\protect\delta _{2}=0.04$, $\protect\delta %
_{3}=0.06$, $\protect\delta _{4}=0.08$; $\protect\zeta _{10}=-0.01$, $%
\protect\zeta _{20}=0.01$, $\protect\zeta _{210}=0.03$, $\protect\zeta %
_{220}=0.04$, $\protect\zeta _{310}=-0.05$, $\protect\zeta _{320}=0.06$, $%
\protect\zeta _{410}=0.07$, $\protect\zeta _{420}=0.08$; $\protect\chi %
_{1}=0.01$, $\protect\chi _{2}=0.03$, $\protect\chi _{3}=0.05$, $\protect%
\chi _{4}=0.07$; $\protect\varepsilon _{1}^{(1)}=0.4$, $\protect\varepsilon %
_{1}^{(2)}=0.2$, $\protect\varepsilon _{1}^{(3)}=0.3$, $\protect\varepsilon %
_{1}^{(4)}=0.2$, with time-dependent functions $\Gamma (t)=-0.035t$ and $%
\protect\alpha (t)=0.1\cos (0.2t)$.}
\end{figure}
\begin{figure}[tbp]
\begin{center}
\includegraphics[width=0.45\linewidth]{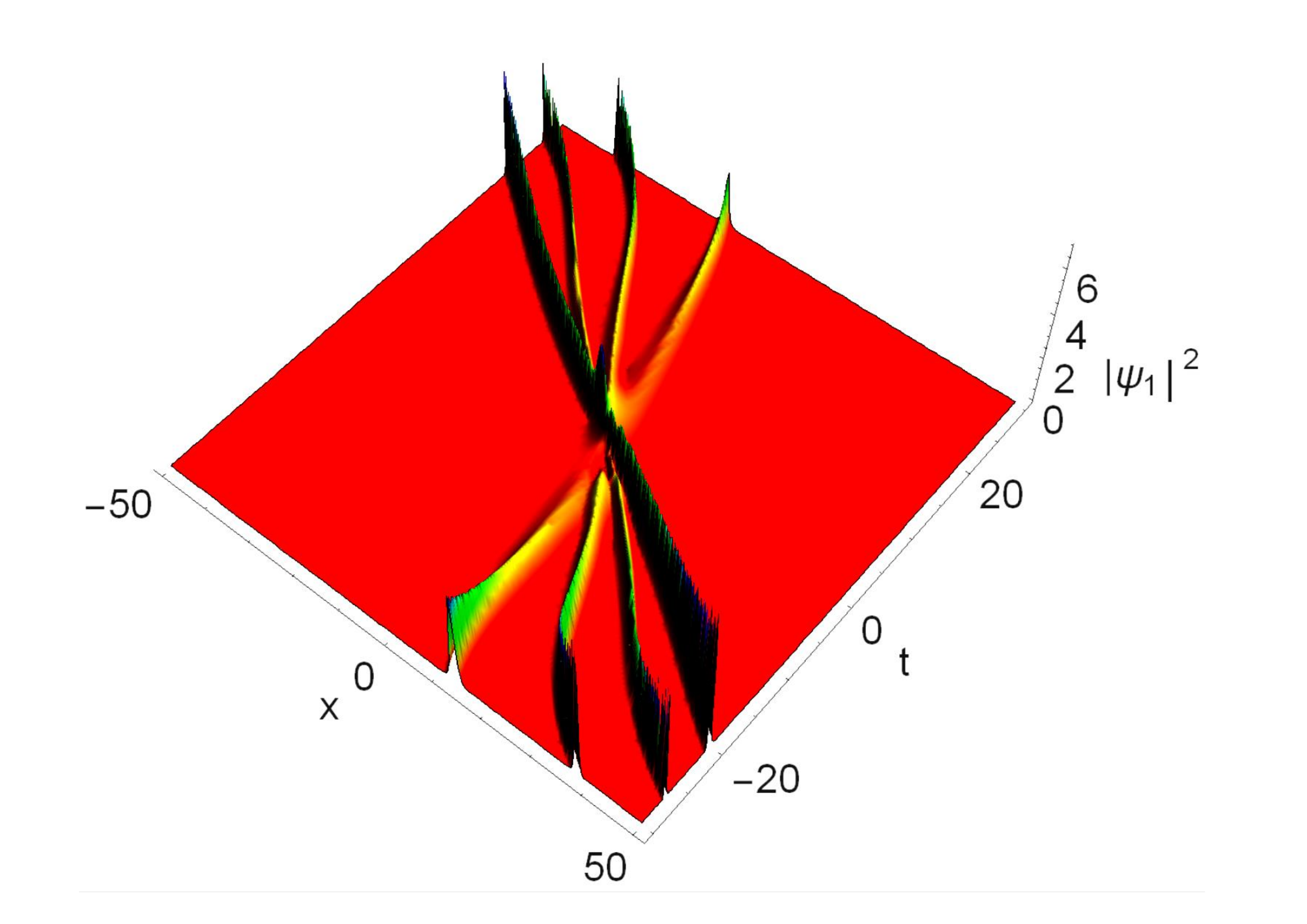} %
\includegraphics[width=0.45 \linewidth]{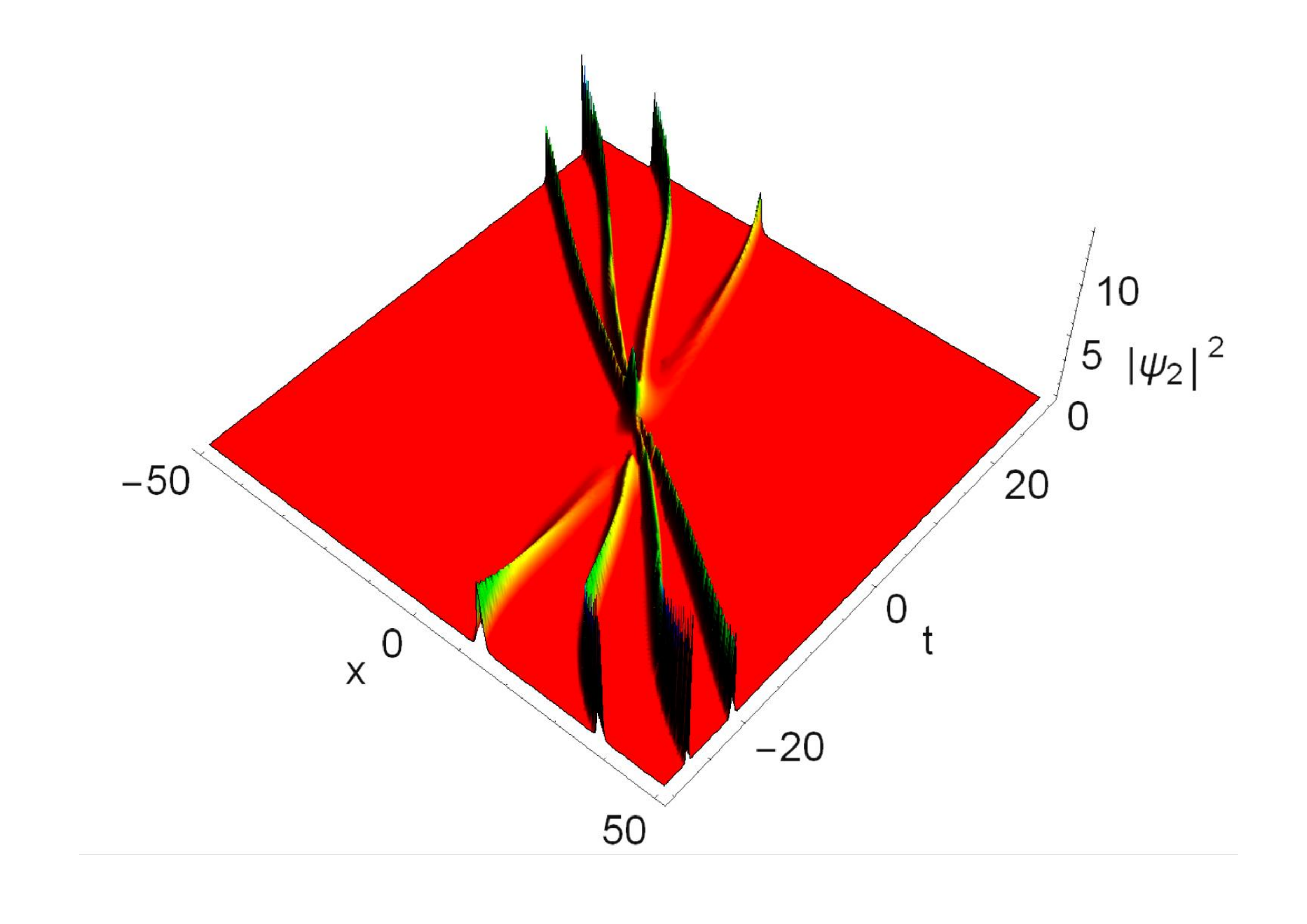}
\end{center}
\caption{ Elastic collision of four vector solitons governed by Eq.~(39)
under a time-dependent trap and nonlinearity. The soliton amplitudes are
controlled by parameters $\protect\delta _{1}=0.75$, $\protect\delta %
_{2}=0.82$, $\protect\delta _{3}=0.69$, and $\protect\delta _{4}=0.85$,
while the velocities are set via $\protect\zeta _{10}=0.1$, $\protect\zeta %
_{20}=0.2$, $\protect\zeta _{210}=0.3$, $\protect\zeta _{220}=0.4$, $\protect%
\zeta _{310}=0.5$, $\protect\zeta _{320}=0.6$, $\protect\zeta _{410}=0.7$,
and $\protect\zeta _{420}=0.8$. The phase offsets are given by $\protect\chi %
_{1}=0.9$, $\protect\chi _{2}=0.75$, $\protect\chi _{3}=0.8$, and $\protect%
\chi _{4}=0.9$, and the polarization distribution is defined by $\protect%
\varepsilon _{1}^{(1)}=0.6$, $\protect\varepsilon _{1}^{(2)}=0.5$, $\protect%
\varepsilon _{1}^{(3)}=0.4$, and $\protect\varepsilon _{1}^{(4)}=0.7$. The
external modulation is imposed by $\Gamma (t)=-0.002t$ and $\protect\alpha %
(t)=0.1\exp (0.02t)$. The solitons exhibit elastic interactions, preserving
their shape and velocity after collision. }
\end{figure}
\textbf{All analytical figures were plotted using the exact four-soliton
solution Eq. (39) derived in section 4. Figure 1 was generated using the
split-step Fourier method applied to }Eqs.\textbf{\ (3)-(4) with the
time-dependent coefficients.} An important feature of this system is its
ability to maintain coherence through controlled oscillations of $\alpha (t)$%
. This feature prevents the onset of the decoherence, which is a major
challenge in quantum computing. The ability to fine-tune soliton
interactions via velocity and position parameters allows one to maintain
precise control over quantum-gate operations, ensuring robust and
fault-tolerant quantum computations. \textbf{Compared to earlier studies on
two-soliton interactions \cite{vrk2010}, the current four-soliton setting
exhibits higher-order collective behavior such as simultaneous fission and
elastic scattering. Unlike the binary soliton logic suggested in \cite%
{tvngo2021}, our approach enables more complex superposition-controlled
fission and symmetric energy splitting, as illustrated in Figures 2-4. These
processes are facilitated by the extra degrees of freedom introduced by
additional spectral parameters $\zeta _{j0}$, amplitude controls $\delta _{j}
$, and polarization coefficients $\varepsilon _{j}$. The ability to tune
these parameters independently allows fine-tuned control over the soliton
trajectories, phase evolution, and population distribution between the
components. This makes the four-soliton setting a promising platform for
simulating more complex quantum-like dynamics.} \\

Thus, the controlled fission
of the solitons in the two-component BEC has profound implications for the
qubit control in quantum computing. In that context, qubits are fundamental
data units, capable to exist in superpositions of states. Matter-wave
solitons in BEC can represent localized quantum states, due to their
stability and coherence, making them potential candidates for qubit encoding
and manipulations. The controlled fission of solitons, when a multi-soliton
state splits into separate solitons, is similar to the quantum-gate
operation, in which an entangled qubit system undergoes a controlled
separation or interaction. \textbf{This extension from two- to four-soliton
states enriches the soliton-based qubit modeling, previously addressed in
work \cite{tvngo2021}, by enabling enhanced state encoding via the
polarization and velocity control. } We stress that the displayed figures
represent typical dynamical behaviors of multi-soliton states under a broad
range of parameter choices. While specific values were selected to highlight
key features, such as controlled fission, superposition, and elastic
collisions, these outcomes are robust, persisting across variations in
soliton amplitudes, phases, and interaction coefficients, and thus
indicating that the results reflect generic situations in the framework of
the integrable model. \textbf{We note that the model's integrability is
predicated on the constraints pertaining to the intra- and interatomic
interaction in the presence of the external trapping potential, and that
gain is hard to implement in practical BEC experiments. Nevertheless,
integrable models can be used as accurate approximations for finite-time
evolution or under strict experimental control, offering valuable insight
into the picture of soliton interactions.}\\

\textbf{The key parameters
governing the soliton dynamics in the present context are summarized as
follows: $\zeta _{j0}$ controls the velocity of the $j$-th soliton, and $%
\delta _{j}$ determines the amplitude and energy distribution. Parameter $%
\chi _{j}$ represents the relative phase between the solitons, and $%
\varepsilon _{j}$ describes the population distribution between the two
components, similar to coefficients of a qubit state. The time-dependent
trap parameter $\Gamma (t)$ affects the soliton's chirp, spatial separation,
and effective phase evolution. In Figures~2--4, parameters such as $\zeta
_{j0}$ determine the soliton's velocity via relation $v_{j}=-2\zeta
_{j0}\,\exp \left( -2\int \Gamma (t)\,dt\right) $, while $\delta _{j}$
controls the amplitude scaling factor. The choice of these parameters
directly affects the separation degree, overlap, and interference between
the solitons. }

\section{Conclusion}

In this work, the dynamics of vector solitons in the two-component BRC\
(Bose-Einstein condensate) governed by coupled Gross-Pitaevskii equations
were systematically investigated. Explicit multi-soliton solutions,
including two-, three-, and four-soliton states, were derived using the
gauge-transformation method. We have highlighted the effect of the
time-dependent intra- and inter-component interaction coefficients and
external potential on the soliton's amplitude and its evolution. The
controlled four-soliton fission was demonstrated, showing the ability to
manipulate the soliton separation and interaction by means of the external
parameters. We have also identified elastic four-solitons collisions, with
the solitons retaining their identity post-interaction. These findings
underscore the potential of the two-component solitons for applications to
coherent matter-wave transport, nonlinear optics, and quantum information
processing -- in particular, for modeling qubit states and gate operations.
Future research may uphold experimental realizations of these soliton states
and their use in quantum technologies.

\section*{Acknowledgments}

The work of B.A.M. was supported, in part, by the Israel Science Foundation
through grant No. 1695/22.

\end{document}